\documentclass[preprint,12pt]{elsarticle}

\usepackage{amscd,amsmath,a4,amsfonts,amssymb,amsthm,enumerate,mathdots}

\usepackage{hyperref}
\usepackage{cleveref}
\usepackage{mathrsfs}
\usepackage{ dsfont }
\usepackage{graphicx}
\usepackage{float}
\usepackage{bbm}
\usepackage{comment}
\usepackage{hyperref}
\usepackage{listings}
\usepackage{color}
\usepackage[dvipsnames]{xcolor}
\usepackage[lined,boxed,commentsnumbered]{algorithm2e}
\usepackage[colorinlistoftodos]{todonotes}
\usepackage[dvipsnames]{xcolor}
\usepackage{tabularx}
\usepackage{comment}
\usepackage{ulem}

\DeclareMathOperator*{\argmin}{argmin}

\begin{document}
\begin{frontmatter}

\title{Data-driven Identification of Parametric Governing Equations of Dynamical Systems Using the Signed Cumulative Distribution Transform}

\author[inst1,inst3]{Abu Hasnat Mohammad Rubaiyat}

\affiliation[inst1]{organization={Department of Electrical and Computer Engineering},
            addressline={University of Virginia}, 
            city={Charlottesville},
            state={VA},
            postcode={22904}, 
            country={USA}}

\author[inst2]{Duy H.~Thai}

\affiliation[inst2]{organization={Department of Biomedical Engineering},
            addressline={University of Virginia}, 
            city={Charlottesville}, 
            state={VA},
            postcode={22908},
            country={USA}}

\author[inst3]{Jonathan M.~Nichols}
\author[inst6]{Meredith N.~Hutchinson}

\affiliation[inst3]{organization={U.S. Naval Research Laboratory},
            city={Washington},
            state={DC},
            postcode={20375}, 
            country={USA}}

\author[inst4]{Samuel P.~Wallen}
\author[inst4]{Christina J.~Naify}
\author[inst4]{Nathan Geib}
\author[inst5,inst4]{Michael R.~Haberman}

\affiliation[inst4]{organization={Applied Research Laboratories},
            addressline={The University of Texas at Austin}, 
            city={Austin}, 
            state={TX},
            postcode={78712},
            country={USA}}

\affiliation[inst5]{organization={Walker Department of Mechanical Engineering},
            addressline={The University of Texas at Austin}, 
            city={Austin}, 
            state={TX},
            postcode={78712},
            country={USA}}

\affiliation[inst6]{organization={Office of Naval Research},
            city={Arlington},
            state={VA},
            postcode={22217}, 
            country={USA}}

\author[inst2,inst1]{Gustavo K.~Rohde}

\begin{abstract}
This paper presents a novel data-driven approach to identify partial differential equation (PDE) parameters of a dynamical system.
Specifically, we adopt a mathematical ``transport'' model for the solution of the dynamical system at specific spatial locations that allows us to accurately estimate the model parameters, including those associated with structural damage. This is accomplished by means of a newly-developed mathematical transform, the signed cumulative distribution transform (SCDT), which is shown to convert the general nonlinear parameter estimation problem into a simple linear regression. This approach has the additional practical advantage of requiring no {\it a priori} knowledge of the source of the excitation (or, alternatively, the initial conditions). By using training data, we devise a coarse regression procedure to recover different PDE parameters from the PDE solution measured at a single location. Numerical experiments show that the proposed regression procedure is capable of detecting and estimating PDE parameters with superior accuracy compared to a number of recently developed machine learning methods. Furthermore, a damage identification experiment conducted on a publicly available dataset provides strong evidence of the proposed method's effectiveness in structural health monitoring (SHM) applications.
The Python implementation of the proposed system identification technique is integrated as a part of the software package PyTransKit \cite{pytranskit}.
\end{abstract}

\begin{keyword}
Structural health monitoring, system identification, signed cumulative distribution transform, elastic waves, ultrasonics
\end{keyword}

\end{frontmatter}

\section{Introduction} \label{sec:intro}

Discovering information regarding the governing equations of a dynamical system is an important step in the design of solutions to many important modern problems in science and technology. The idea is to discover a model (e.g., the parameters comprising the coefficients of a differential equation) from existing data for the purposes of understanding the intricacies of the phenomenon at hand, as well as to perform useful predictions about the future behavior of the system. Applications are plentiful, ranging from time series prediction (e.g., finance), weather modeling, modeling signaling networks in biology, and others. One example application is the field of Structural Health Monitoring (SHM) where the goal is to use measured data to assess the condition of a structure \cite{guemes2020structural}. A critical step is the identification and classification of parameters of the underlying governing equations (PDE), which in turn can be related to the structure's ``health'' i.e., the structural integrity \cite{noel2017nonlinear,an2019recent}.  Typically, it is presumed that this information is to be inferred from the dynamic response of the structure to ambient or applied excitation, i.e. a measured acoustic signal \cite{Nichols:16}. 

While numerous approaches to the problem of PDE system identification exist in many different settings or applications, they can be loosely categorized in terms of the {\it a priori} information required for their implementation. On one hand, we can view the problem as one of ``statistical pattern recognition'' \cite{Worden:07} whereby certain properties of the response data (referred to in the literature as ``features'') are used to detect and then classify the response as coming from a particular state of the structure, e.g., nonlinearities, dispersion, etc. Recent advances in machine learning have furthered research in this general approach \cite{Worden:12,noel2017nonlinear}. NARMAX (Non-linear AutoRegressive Moving Average with eXogenous inputs) \cite{chen1989representations}, neural networks \cite{gonzalez1998identification,takeishi2017learning, wehmeyer2018time, yeung2019learning}, equation-free methods \cite{kevrekidis2003equation, kevrekidis2009equation}, and Laplacian spectral analysis \cite{giannakis2012nonlinear} are among other data-driven methods that have been used in system identification problems. Sparse regression-based methods have also been studied in recent literature to identify ordinary \cite{brunton2016discovering, chen2017network} and partial \cite{rudy2017data, schaeffer2017learning, rudy2019data} differential equations. However, these methods often require significant amounts of data for training, involve tuning a large number of parameters, and can be computationally expensive.
Another key challenge with the associated data models lies in capturing the target aspects of the measured response data while ignoring those due to covariates (e.g., measurement noise, temperature variations, etc.) \cite{Sohn:02}. A model that is unable to distinguish among these sources will produce an unacceptably high number of errors (e.g., ``false positives'') mis-categorizing other influences on the system response as being a specific condition of the system.

Another class of techniques views the problem as one of model-based system identification \cite{Nichols:16,rudy2019data}. In this setting, a model of the system is formulated and various estimation methods are used to identify the parameters related to structural damage.  One of the main advantages of such an approach is at least partial immunity to data fluctuations un-related to parameters of interest. The model is explicitly separating the physics of interest (e.g., damage in structural health monitoring) from these ``other'' sources which are collectively modeled as noise, confounds, or nuisances. The challenge with this approach lies in the modeling. Predicting the response of a physical system to a particular excitation is challenging, as the exact forcing function in a given application is often unknown (particularly if ambient excitation is taken as the forcing, e.g., wind or waves) \cite{Nichols:12}.

These different viewpoints (data driven vs. model based) are not mutually exclusive and represent endpoints on the continuum of {\it a priori} information we wish to bring to bear on the problem. 
The method proposed in this work lies between these aforementioned extremes. We propose a PDE coefficient estimation approach that utilizes \textit{a priori} training data, as in a time series (signal) classification task. When the general form of the PDE is known, training data can be obtained via simulation using randonly chosen coefficients as labels. When the PDE is unknown, as is the case for the majority of structural health monitoring applications, experimental training data can be used. In addition, we assume the system identifcation problem is well-posed in that the state of the system (e.g. parameter of interest) is in one-to-one correspondence in the measurements.  However, we do not presume to know or be able to measure the excitation signal.

\subsection{Motivating example and overview of approach:}\label{sec:motiv_example}

\begin{figure}[tb]
    \centering
    \includegraphics[width=0.9\textwidth]{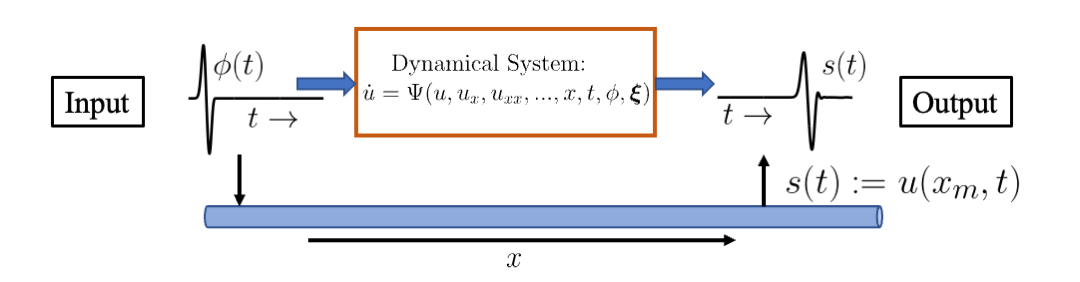}
    \caption{A model dynamical system describing 1D wave propagation in an elastic domain.}
    \label{fig:1d_dyn_sys}
    \vspace{-1.0em}
\end{figure}

Consider a model dynamical system (depicted in Fig.~\ref{fig:1d_dyn_sys}) describing a 1D wave propagation phenomenon through an elastic medium. Typically, the dynamics of such systems are expressed as partial differential equations (PDEs) \cite{rudy2019data} in the following form:
\begin{equation}
    \dot{u} = \Psi(u,u_{x},u_{xx},...,x,t, \phi, \boldsymbol{\xi}),
    \label{eq:sys_model}
\end{equation}
where $\dot{u}=\frac{\partial u}{\partial t}$, $u_x=\frac{\partial u}{\partial x}$, and $u_{xx}= \frac{\partial^2 u}{\partial x^2}$ and so on. $\Psi$ represents a PDE-based model for the dynamical system that is parameterized with $\boldsymbol{\xi}$. $\phi(t)$ denotes the source signal that initiates the propagation and $s(t) = u(x_m,t)$ is the sensor measurement measured at a specific location ($x=x_m$) as a function of time as depicted in Fig.~\ref{fig:1d_dyn_sys}. The main objective of system identification problems is to estimate some property of the system (e.g. the system coefficients $\boldsymbol{\xi}$ of the governing PDE) using the sensor measurement $s(t)$. In this work, the focus is primarily on identifying the coefficients associated with damage-induced nonlinearities, though the approach is general and can be used to estimate any PDE coefficient under the assumptions that 1) each $\boldsymbol{\xi}$ is associated with a unique output $s(t)$ and 2) $s(t)$ varies smoothly as a function of the particular coefficient of interest.

Let us begin with a simple example of a wave propagating in the elastic rod depicted in Fig.~\ref{fig:1d_dyn_sys}. For a linear elastic material in the absence of dispersion and loss, wave motion is described by the model \cite{Achenbach:73}
\begin{equation}
    \ddot{u} -  \nu^2 u_{xx} = \phi.
    \label{eq:lin_we_source}
\end{equation}
Here $u(x,t)$ refers to the local longitudinal wave displacement \cite{Achenbach:73}, $\nu$ is the speed of propagation of a disturbance in the medium (the phase speed), and the source function $\phi(t)$, applied at a particular location, initiates the longitudinal wave along the propagation medium. Thus, the generated displacement $u(x,t)$ is a response to $\phi(t)$. As shown in \cite{gazdag1981modeling}, for example, an equivalent way of stating the problem is to define a proper set of initial conditions, e.g., defining $u$ and $\dot{u}$ at $t=0$, and setting the source function $\phi$ to be zero. Using this approach, we can rewrite Eq.~(\ref{eq:lin_we_source}) as:
\begin{equation}
    \ddot{u} -  \nu^2 u_{xx} = 0,
    \label{eq:lin_we_wo_source}
\end{equation}
which can be separated into two PDEs of the 1st order using the method of PDE factorization \cite{lamoureux2006mathematics}:
\begin{align}
    &\dot{u} - \nu u_x = 0 \label{eq:lin_we_neg_x} \\
    &\dot{u} + \nu u_x = 0. \label{eq:lin_we_pos_x}
\end{align}

Let us consider the PDE defined in Eq.~(\ref{eq:lin_we_pos_x}), which describes the wave propagation along $+x$ direction.
In this case the solution (our measured signal at location $x$ in the rod) is $s_\nu(t) = u(x,t) = \varphi(t - x/\nu)$, for some arbitrary initial condition $\varphi$ \cite{Achenbach:73}. Note that the notation $s_{\nu}(t)$ specifies that the measured time domain acoustic signal is a function of the PDE parameter $\nu$ (wave speed).
We now make the substitution $g(t,x) = t - x/\nu$ and note that $\dot{g}(t,x) = 1$. Therefore, we can write the signal measured at some location $x=x_m$ in the rod as 
\begin{equation}
    s_{\nu}(t) = \dot{g}(t,x_m)\varphi(g(t,x_m)),
    \label{eqn:transportJMN}
\end{equation}
where $\varphi$ is a template function, which, in this case, is the arbitrary initial condition.
Eq.~(\ref{eqn:transportJMN}) is a very general model of PDE's solution at location $x_m$ (i.e., response data), stating that the measured data will be a deformed version of $\varphi$ under the action of a function $g(t,x_m)$ governing the dynamics. The model also implies that the total amplitude $\int\varphi dt$ is preserved during propagation along a lossless medium. Such a model has motivated data estimation and classification approaches based on the cumulative distribution transform (CDT) \cite{park2018cumulative} and the signed cumulative distribution transform (SCDT) \cite{aldroubi2022signed}, which are recently introduced mathematical transforms that are reviewed below. Using the CDT, we can rewrite the equation above in CDT space as:
\begin{equation}
    \widehat{s}_\nu(y) = g^{-1}( \widehat{\varphi}(y),x_m) = \widehat{\varphi}(y) + x_m/\nu,
    \label{eq:cdt_example_translation}
\end{equation}
where the hat symbol indicates the function expressed in transform domain $y$ and with $g^{-1}(t,x_m) = t + x_m/\nu$. That is $g^{-1}(g(t,x_m),x_m) = t$ is the functional inverse of $g(t,x_m)$ along variable $t$. Note that with knowledge of the location $x_m$ at which the measurement $s_\nu$ is taken, and knowledge of the template $\varphi$ (in this case the response to $\phi$ at source location) we are able to recover the value of $\nu$ via:
\begin{equation}
    \nu  = \frac{x_m}{\int_{\Omega_0}(\widehat{s}_\nu(y) - \widehat{\varphi}(y))dy}
\end{equation}
with $\Omega_0$ denoting the domain of the reference function used in the definition of the CDT (see below and \cite{rubaiyat2020parametric} for more details). 

\begin{figure}[tb]
    \centering
    \includegraphics[width=0.95\textwidth]{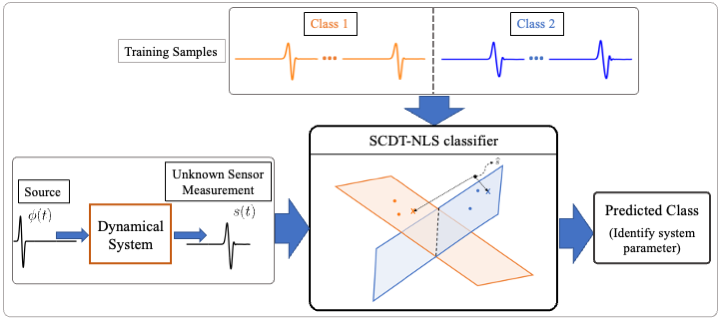}
    \caption{Overview of the system identification approach described in this work.}
    \label{fig:overview}
    \vspace{-1.0em}
\end{figure}

In summary, with knowledge of the template function $\varphi$, as well as the location of the measurement $x_m$, we are able to recover the parameter of the transport PDE (\ref{eqn:transportJMN}) via {\it a simple linear regression in the transform domain}. Below we generalize the procedure to more general PDEs, where the template function $\varphi$ may not be known, and show that under appropriate smoothness assumptions, the CDT technique can linearise such estimation problems locally (within a small range for the parameter of interest). By using training data, which can be obtained from either computer simulations or real lab-acquired measurements, we describe a regression procedure capable of retrieving accurate information regarding the state of the system, and even measurements of different PDE parameters from one single time-domain acoustic sensor, as depicted in Fig.~\ref{fig:1d_dyn_sys}. Importantly, we note that our proposed regression procedure does not require knowledge of the initial condition $\varphi$, rather it only uses simulated or measured data. Thus, the prescribed approach removes a major obstacle to model-based system identification (see e.g., section 5.2 of Ref.~\cite{Nichols:12}). Fig.~\ref{fig:overview} demonstrates the overview of the proposed regression-based system identification procedure.

This paper is organized as follows. In section \ref{sec:preli}, we define the mathematical notation and abbreviations utilized throughout this paper. This includes a definitions of the cumulative distribution transform (CDT), the signed cumulative distribution transform (SCDT), and the nearest local subspace (NLS) method for data classification, all of which are existing mathematical methods for signal processing and data analysis from prior work. In section \ref{sec:model_sysId}, we then describe how one can use these tools to estimate the coefficients of the standard convection diffusion equation, for which the analytical solution is known, using a standard training (calibration) and testing (regression) signal classification approach.
In section \ref{sec:method}, we define the system identification problem related to estimation of PDE coefficients from a more general model and describe our simulation and regression method-based estimation procedure. The proposed method employs the SCDT and NLS methods reviewed in section \ref{sec:preli}. Both the training (calibration) and testing (estimation) phase are described in detail. In section \ref{sec:experiments}, we describe our numerical implementation of the method as well as report the validation experiments we have performed for PDE parameter estimation. We also describe regression (classification) methods based on Deep Learning to which we compare our regression results. In addition to the numerical validations, we have performed a damage identification task on a publicly available dataset to demonstrate the efficacy of our proposed method in the context of structural health monitoring (SHM). The outcomes of this experiment are detailed in section \ref{sec:data_experiment}. Concluding remarks are provided in section \ref{sec:conclusion}.

\section{Preliminaries}\label{sec:preli}
\subsection{Notation}
In the following sub-sections, we introduce the CDT, SCDT and SCDT-NLS for signals $s \in L_1(\Omega_s)$ on the signal domain $\Omega_s \subseteq \mathbb R$. We denote $s^{(c)}_{j}$ as a signal generated by a warping map $g_j$ acting on template $\varphi^{(c)}$ of class $c$. The function $g_j$ captures the influence of the structure (including the damage) on the input $\varphi^{(c)}$ for generic structural systems supporting the basic physics of wave propagation as captured by Eq.~(\ref{eqn:transportJMN}). In addition, we at times make use of the common notation: $\| s \|^2 = <s, s> = \int_{\Omega_s} s(x)^* s(x) dx = \int_{\Omega_s} |s(x)|^2 dx $, where $<\cdot, \cdot>$ is the inner product. Signals are assumed to be real, so the complex conjugate $^*$ does not play a role. Some other notations used throughout the manuscript are listed in Table \ref{tab:symbols}. 

\begin{table}[bt]
    \centering
    \normalsize
    \caption{Description of symbols}
        \begin{tabular}{ll}
        \hline
        Symbols                & Description    \\ \hline
        $\phi(t)$ & Source signal that excites the propagation medium \\
        $s(t)$ & Sensor measurement as a function of time \\
        $s_0(y)$ & Reference signal to calculate the transform\\
        $\widehat{s}(y)$ & SCDT of signal $s(t)$\\
        $\varphi(t)$ & Template pattern corresponding to a class of signals\\
        $g(t)$ & Strictly increasing and differentiable function \\
        $s\circ g$ & $s(g(t))$: composition of $s(t)$ with $g(t)$\\
        $\mathcal{T}$ & Set of all possible increasing diffeomorphisms\\
        $\mathbb{S}/\widehat{\mathbb{S}}$ & Set of signals$/$SCDT of the signals \\
        \hline
        \end{tabular}
    \label{tab:symbols}
\end{table}

\subsection{The Cumulative Distribution Transform} \label{sec:CDT}

The CDT \cite{park2018cumulative} of positive smooth normalized functions is an invertible nonlinear 1D signal transform from the space of smooth positive probability densities to the space of diffeomorphisms.
Given, a signal $s(t), t\hspace{2pt}\in\Omega_s$ and a reference signal $s_0(y),y\in\Omega_{s_0} \subseteq \mathbb R$ such that $\int_{\Omega_s}s(u)du = \int_{\Omega_{s_0}}s_0(u)du = 1$ and $s_0(y), s(t)>0$ in their respective domains.
The CDT of the signal $s(t)$ is the function $s^*(y)$ computed as:
\begin{align}
    \int_{\inf(\Omega_s)}^{s^*(y)} s(u)du = \int_{\inf(\Omega_{s_0})}^{y} s_0(u)du ,
    \label{eq:cdt}
\end{align}
%
%
which can alternatively be defined as:
\begin{equation}
	s^*(y) = S^{-1}(S_0(y)) \,,
	\label{eq:cdt_alt}
\end{equation}
where $S(t) = \int_{-\infty}^{t} s(u)du$ and $S_0(y) = \int_{-\infty}^{y} s_0(u)du$. 
If the reference signal is uniform, i.e.~$s_0(y) = 1$ in $\Omega_{s_0}=[0,1]$, we have $S_0(y) = y$ and therefore, $s^*(y) = S^{-1}(y)$. 

Although the CDT can widely be used in classification \cite{park2018cumulative} and estimation \cite{rubaiyat2020parametric} problems, it is defined only for positive density functions.
Aldroubi et al.~\cite{aldroubi2022signed} proposed the signed cumulative distribution transform (SCDT) as an extension of the CDT to general finite signed signals.

\subsection{The Signed Cumulative Distribution Transform} \label{sec:SCDT}

\begin{figure}[tb]
    \centering
    \includegraphics[width=0.7\textwidth]{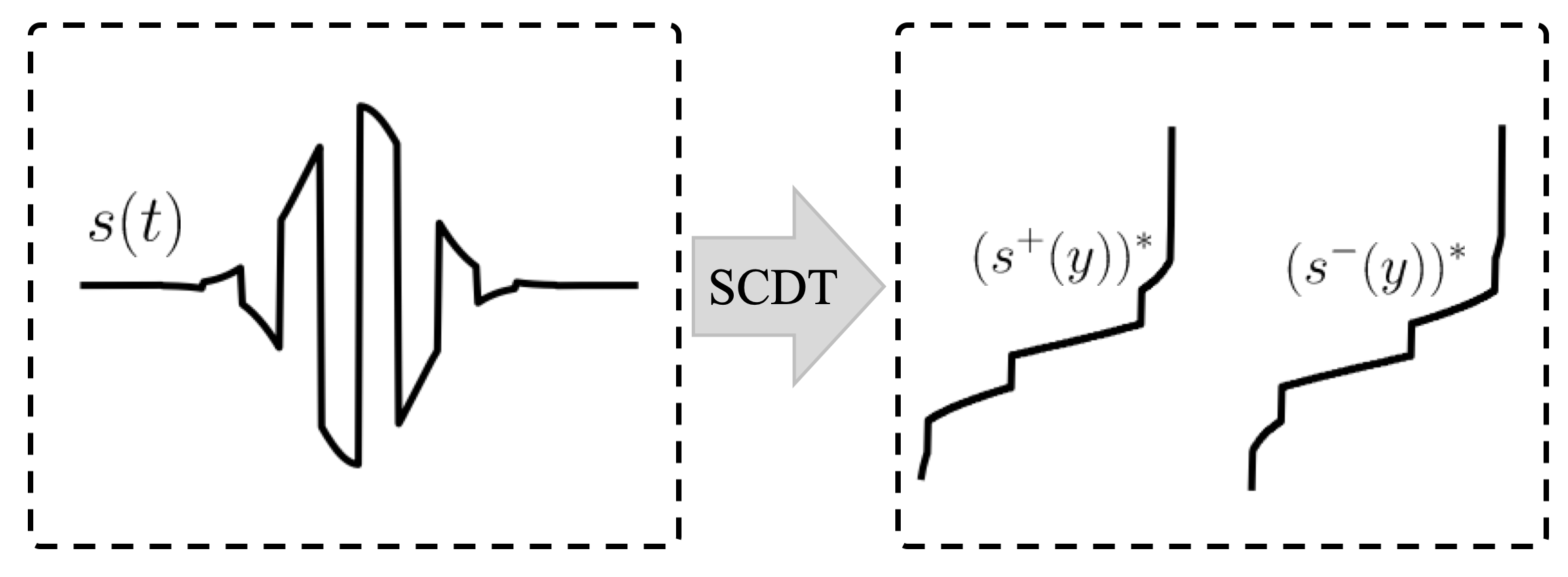}
    \caption{SCDT (without the constant terms) of an example signal. 
    }
    \label{fig:scdt_ch4}
    \vspace{-1em}
\end{figure}

The SCDT \cite{aldroubi2022signed} is an extension of the CDT for general finite signed signals without requirements on the total ``mass" (signal intensity).
Given a signed signal $s(t)$, the Jordan decomposition of the signal is given by $s(t) = s^+(t) - s^-(t)$, where $s^+(t)$ and $s^-(t)$ are the absolute values of the positive and negative parts of $s(t)$, respectively. The SCDT of $s(t)$ with respect to $s_0(y)$ is then defined as:
\begin{equation}
 s(t)
 ~\stackrel{\text{SCDT}(s_0)}{\longleftrightarrow}~
 \widehat{s}(y) = \left(\widehat{s}^+(y), \widehat{s}^-(y)\right),
 \label{eq:scdt}
\end{equation}
where $\widehat{s}^+(y)$ and $\widehat{s}^-(y)$ are the transforms for the signals $s^+(t)$ and $s^-(t)$ as:
\begin{equation}
    \widehat{s}^\pm(y) = \begin{cases}
    \left(\left(s^\pm\right)^*(y),\|s^\pm\|_{L_1}\right),& \text{if } s\neq 0\\
    (0,0),              & \text{if } s=0,
\end{cases}
\label{eq:scdt_mass}
\end{equation}
with $L_1$ norm $\|\cdot\|_{L_1}$ and $\left( s^\pm \right)^*$ as the CDT of the normalized signal $\frac{s^\pm}{\|s^\pm\|_{L_1}}$ with respect to a strictly positive reference signal $s_0$. Fig.~\ref{fig:scdt_ch4} demonstrates the SCDT calculation of an example signal.
%
%
The SCDT has a number of properties that are useful for the signal classification problems.

\subsubsection{Composition property}
Composition property states that the SCDT of the signal $s_g=g's\circ g$, is defined as:
\begin{equation}
    \widehat{s}_g = \left(g^{-1}\circ (s^+)^*,\|s^{+}\|_{L_1},g^{-1}\circ (s^-)^*,\|s^{-}\|_{L_1}\right),
    \label{eq:scdt_composition}
\end{equation}
where $g(t)$ is an invertible smooth warping map, $s\circ g = s(g(t))$ and $g'(t)=dg(t)/dt$ \cite{aldroubi2022signed}.
For example, a shift and linear dispersion (i.e., $g(t) = \omega t - \tau$) of a given signal $s(t)$ is 
$s_g(t)=\omega s(\omega t - \mu)$
having SCDT:
\begin{equation*}
    \widehat{s}_g = \left(\frac{(s^+)^* + \mu}{\omega},\|s^{+}\|_{L_1},\frac{(s^-)^* + \mu}{\omega},\|s^{-}\|_{L_1}\right).
\end{equation*}
The composition property implies that variations along the independent variable caused by $g(t)$ will change only the dependent variable in the transform domain. 

\subsubsection{Convexity property}
\label{sec:convexity}
Given a signal $\varphi$ and 
a set of 1D temporal deformations $\mathcal{G}$ (e.g., translation, dilation, etc), 
the set of the SCDTs of the signals from 
$\mathbb{S}=\{s_j : s_j=g'_j \varphi\circ g_j, \forall g_j\in \mathcal{G}\}$
is given by   
$\widehat{\mathbb{S}} = \{\widehat{s}_j:\widehat{s}_j=g_j^{-1}\circ \widehat{\varphi}, s_j\in \mathbb{S}\}$ (using the composition property).
The convexity property of the SCDT \cite{aldroubi2022signed} states that the set $\widehat{\mathbb{S}}$ is convex for every $\varphi$ if and only if $\mathcal{G}^{-1}=\{g_j^{-1}:g_j\in \mathcal{G}\}$ is convex.
The set $\mathbb{S}$ defined above can be interpreted as a mathematical model for a signal class while $\varphi$ being the template signal corresponding to that class. 

\subsubsection{Numerical Implementation of the SCDT}
The SCDT described above is defined for the continuous-time signals. In this section, we describe the numerical method for approximating the SCDT given discrete signals for a particular choice of reference signal ($s_0(y)=1$ for $y\in [0,1]$). Let $\mathbf{s}=\left[s_1, s_2, ..., s_N \right]^T$ be a $N$-point discrete-time signal, where $s_n=s[n],\forall n=1,2,...,N$ is the $n$-th sample of $\mathbf{s}$. The positive and negative parts of the signal after Jordan decomposition are given by $\mathbf{s}^\pm = \left[s^\pm_1,..., s^\pm_N \right]^T$, respectively, where $s^\pm_n=\frac{|s_n| \pm s_n}{2}$, and $|s_n|$ is the absolute value of $s_n$. Next, the CDT is applied numerically to the normalized signals $\frac{\mathbf{s}^+}{\|\mathbf{s}^+\|_{\ell_1}}$ and $\frac{\mathbf{s}^-}{\|\mathbf{s}^-\|_{\ell_1}}$, where $\|\cdot\|_{\ell_1}$ is the $\ell_1$-norm. As the CDT $(s^\pm)^*(y)$ is the inverse of the cumulation of the signal $s^\pm(t)$, we must first approximate the cumulative function which is numerically given by,
\begin{equation*}
    S^{\pm}[n] = \sum_{i=1}^n \frac{s^\pm[n]}{\|\mathbf{s}^\pm\|_{\ell_1}}, \qquad n=1,2,...,N,
\end{equation*}
where $\mathbf{S}^+$ and $\mathbf{S}^-$ are the cumulation of the normalized signals $\frac{\mathbf{s}^+}{\|\mathbf{s}^+\|_{\ell_1}}$ and $\frac{\mathbf{s}^-}{\|\mathbf{s}^-\|_{\ell_1}}$, respectively. The CDT is then calculated by taking the generalized inverse of $\mathbf{S}^\pm$,
\begin{align*}
    (s^\pm)^*[m]=\min\left(\{t[n]:S^\pm[n]>y[m]\} \right),
\end{align*}
where $t\in \Omega_s$, $y\in\Omega_{s_0}$, and $n,m=1,2,...,N$. Here, $(\mathbf{s}^+)^*$ and $(\mathbf{s}^-)^*$ are the CDTs of the normalized discrete signals $\frac{\mathbf{s}^+}{\|\mathbf{s}^+\|_{\ell_1}}$ and $\frac{\mathbf{s}^-}{\|\mathbf{s}^-\|_{\ell_1}}$, respectively. The SCDT of the discrete signal $\mathbf{s}$ is then given by,
\begin{equation*}
    \widehat{\mathbf{s}} = \left((\mathbf{s}^+)^*, \|\mathbf{s}^+\|_{\ell_1}, (\mathbf{s}^-)^*, \|\mathbf{s}^-\|_{\ell_1} \right).
\end{equation*}

\subsection{Nearest Local Subspace Classifier in SCDT Domain} \label{sec:scdt-nls}

As described below, this work proposes a regression-based method for system parameter identification and solves it using the SCDT nearest local subspace (SCDT-NLS) classification method. The SCDT-NLS \cite{rubaiyat2022end} was proposed as an end-to-end classification technique to classify finite duration signal classes that can be seen as observations of a set of template patterns that have undergone some temporal deformations. Such class of signals can be modeled as:
\begin{align}
    &\mathbb{S}^{(c)} = \bigcup\limits_{m=1}^{M_c} \mathbb{S}_{\varphi_m^{(c)},\mathcal{G}_m^{(c)}}, \nonumber\\
    &\mathbb{S}_{\varphi_m^{(c)},\mathcal{G}_m^{(c)}}=\left\{s_{j,m}^{(c)}|s_{j,m}^{(c)}=g'_{j}\varphi_m^{(c)}\circ g_{j}, g'_{j}>0, g_{j}\in \mathcal{G}_m^{(c)}\right\},\nonumber\\
    &\left(\mathcal{G}_m^{(c)}\right)^{-1}=\left\{\sum_{i=1}^k\alpha_i f_{i,m}^{(c)}, \alpha_i\geq 0\right\},
    \label{eq:genmod_multi}
\end{align}
where $c$ denotes a particular class, $\left\{f_{1,m}^{(c)}, f_{2,m}^{(c)},..., f_{k,m}^{(c)}\right\}$ denotes a set of linearly independent and strictly increasing functions, and $\mathcal{G}^{(c)}_m$ denotes a set of increasing 1D temporal deformations. It states that the signal class model for a given class $c$ can be expressed as the union of $M_c$ subsets, where each subset ($\mathbb{S}_{\varphi_m^{(c)},\mathcal{G}_m^{(c)}}$) represents data generated by a specific template signal ($\varphi_m^{(c)}$) under various temporal deformations ($\mathcal{G}_m^{(c)}$). Here, $M_c$ refers to the total number of templates utilized to represent class $c$, while $\varphi_m^{(c)}$ denotes the $m$-th template signal from class $c$, and $s_{j,m}^{(c)}$ refers to a signal generated from the template $\varphi_m^{(c)}$, subjected to the deformation defined by $g_j$.

Since $\left(\mathcal{G}_m^{(c)}\right)^{-1}$ is convex by construction, the set $\mathbb{S}^{(c)}$ can be represented as a collection of convex sets in the SCDT space, as per the convexity property. the signal class model in the transform space is given by:
\begin{align}
    &\widehat{\mathbb{S}}^{(c)} = \bigcup\limits_{m=1}^{M_c} \widehat{\mathbb{S}}_{\varphi_m^{(c)},\mathcal{G}_m^{(c)}}, \nonumber\\
    &\widehat{\mathbb{S}}_{\varphi_m^{(c)},\mathcal{G}_m^{(c)}} = \left\{\widehat{s}_{j,m}^{(c)}| \widehat{s}_{j,m}^{(c)}=g_{j}^{-1}\circ \widehat{\varphi}_m^{(c)}, g_j\in \mathcal{G}_m^{(c)}\right\},
    \label{eq:genmod_scdt}
\end{align}
where $g_{j}^{-1}\circ \widehat{\varphi}_m^{(c)}$ is the SCDT of the signal $g'_j\varphi_m^{(c)}\circ g_j$. Under the assumption that $\mathbb{S}_{\varphi_m^{(c)},\mathcal{G}_m^{(c)}} \cap \mathbb{S}_{\varphi_w^{(p)},\mathcal{G}_w^{(p)}}=\varnothing$ for $c\neq p$, and an unknown sample $s$ is generated according to the signal class model, the unknown class label can be uniquely predicted by solving,
\begin{equation}
    \argmin_c\min_m~d^2\left(\widehat{s}, \widehat{\mathbb{V}}_m^{(c)}\right), 
    \label{eq:min_problem_nls}
\end{equation}
where $\widehat{\mathbb{V}}_m^{(c)} = \text{span}\left(\widehat{\mathbb{S}}_{\varphi_m^{(c)},\mathcal{G}_m^{(c)}}\right)$, and $d\left(\widehat{s},\widehat{\mathbb{V}}^{(c)}_m\right)$ is the Euclidean distance between $\widehat{s}$ and the nearest point in subspace $\widehat{\mathbb{V}}^{(c)}_m$. Given a set of training samples $\left\{s_1^{(c)},...,s_{L_c}^{(c)}\right\}$ $\subset \mathbb{S}^{(c)}$ for class $c$, the unknown class of a test sample $s$ is estimated in two steps:

\textbf{Step 1:} A set of $k$ closest training samples to $\widehat{s}$ from class $c$ are chosen based on the distance between $\widehat{s}$ and the span of each training sample. First, the elements from the set $\{\widehat{s}_{1}^{(c)}, ..., \widehat{s}_{L_c}^{(c)}\}$ are sorted into $\{\widehat{z}_1^{(c)},...,\widehat{z}_{L_c}^{(c)}\} $ such that 
\begin{align}
	d^2(\widehat{s},\widehat{\mathbb{V}}^{(c)}_{z_1})\leq \cdots \leq d^2(\widehat{s},\widehat{\mathbb{V}}^{(c)}_{z_{l}})\leq \cdots, 
	\label{eq:k-distance}
\end{align}
where $\widehat{\mathbb{V}}^{(c)}_{z_l} = \text{span}\left(\{\widehat{z}_{l}^{(c)}\}\right)$. First $k$ elements from the sorted set are chosen to form $\{\widehat{z}_1^{(c)},\cdots,\widehat{z}_k^{(c)}\}$ for $k\leq L_c$, which gives the set of $k$ closest training samples to $\widehat{s}$ from class $c$ in the above sense. This step is repeated for all the signal classes, i.e., $c=1,2,...,$ etc.

\textbf{Step 2:} The set $\{\widehat{z}_1^{(c)},\cdots,\widehat{z}_k^{(c)}\}$ is orthogonalized to obtain the basis vectors $\{b_1^{(c)},b_2^{(c)},...\}$. 
The unknown class of $s$ is then estimated by:
\begin{equation}
    \arg\min_c~ \|\widehat{s} - B^{(c)}B^{(c)^T}\widehat{s}\|^2,
    \label{eq:test_step_nls}
\end{equation}
where $\|\cdot\|$ denotes $L_2$ norm and $B^{(c)}=\left[b_1^{(c)},b_2^{(c)},... \right]$ for $c=1,2,...,$ etc. It should be noted that $B^{(c)}B^{(c)^T}$ is the orthogonal projection matrix onto the space generated by the span of the columns of $B^{(c)}$. 

\section{Transport-based Modeling Approach for System Identification} \label{sec:model_sysId}


The modeling approach employed here for system identification is based on the observation that Eq.~\eqref{eqn:transportJMN} is well suited for representing solutions to traveling wave problems that describe the dynamics of many structural systems, as already demonstrated in section~\ref{sec:motiv_example}. Here we elaborate on this observation using the standard convection-diffusion equation as a test example. More precisely, we show that the solution to this PDE, for the case when the initial condition is a Gaussian pulse, can be expanded as in Eq.~\eqref{eqn:transportJMN}, and that a local linear expansion around specific values for the wave velocity and diffusion parameters can be modeled as a convex set, which forms the basis for the application of the SCDT-NLS signal classification method.

\subsection{Convection-Diffusion Equation} \label{sec:convdiff}

The 1D convection-diffusion equation with an initial condition is given by:
\begin{align} \label{eq:conv_diffusion}
 &\dot{u} = \nu u_x + D u_{xx}, \\
 \text{initial condition: }&u(x,t=0) = \frac{1}{\sqrt{4 \pi}} e^{-\frac{x^2}{4}},\nonumber
\end{align}
where $\nu$ and $D$ denote the wave speed and the diffusion coefficient, respectively. 
A solution to the PDE defined in Eq.~(\ref{eq:conv_diffusion}) measured at location $x=x_m$ can be derived as:
\begin{align}
    s(t) = \frac{1}{\sqrt{4\pi Dt}}e^{-\frac{(x_m - \nu t)^2}{4Dt}},
    \label{eq:sensor_convdiff}
\end{align}
which can be represented using the following mathematical model:
\begin{align}
    s(t) &= \dot{g}_{\nu,D}(t)\varphi_{\nu,D}(g_{\nu,D}(t)),\quad g_{\nu,D}\in\mathcal{G}_{\nu,D},\\
    &x_m,\nu,D>0,\quad t > \frac{x_m}{\nu}.
\end{align}
The template $\varphi_{\nu,D}(t)$ and the warping function $g_{\nu,D}(t)$ are derived as following:
\begin{align}
    &\varphi_{\nu,D}(t) = \frac{1}{\nu} \sqrt{ \frac{4 D}{\pi} }
 \frac{ \left( \left( x_m \nu + 2 D t \right)
 + \sqrt{ \left( x_m \nu + 2 D t \right)^2 - \nu^2 x_m^2 }
 \right)^{\frac{3}{2}} }
 { \left( \left( x_m \nu + 2 D t \right)
 + \sqrt{ \left( x_m \nu + 2 D t \right)^2 - \nu^2 x_m^2 }
 \right)^2 - \nu^2 x_m^2 } 
 e^{-t} \,, \nonumber \\
 &g_{\nu,D}(t) = \frac{(x_m - \nu t)^2}{4 D t},\quad g_{\nu,D}\in\mathcal{G}_{\nu,D}. 
\end{align}
For a detailed derivation, please refer to \ref{app:conv-diff}. Since $g_{\nu,D}(t)$ is a quadratic polynomial function of $t$, it does not have a unique inverse warping map. However, under the condition that $g_{\nu,D}^{-1}(t)$ is monotonically increasing, the inverse of the warping can be set as,
\begin{align} \label{eq:inverwarpingmap:convecdiff}
 g_{\nu,D}^{-1}(t) &=
 \frac{1}{\nu^2} \left( x_m \nu + 2 D t 
 + \sqrt{ \left( x_m \nu + 2 D t \right)^2 - \nu^2 x_m^2 }
 \right),\, g^{-1}_{\nu,D}\in \mathcal{G}_{\nu,D}^{-1} \,.
\end{align}

\subsubsection{Convex approximation of inverse warping maps}

Our objective here is to show that locally, within a small neighborhood of the parameter of interest, the set of solutions to the convection diffusion PDE can be viewed as a linear subspace in SCDT domain. Consider the case when $D \in [D_0-\epsilon,D_0+\epsilon]$, $D>0$. We aim to devise a system identification technique for estimating $\nu$ using the tools described in section \ref{sec:preli} under the condition that the family of functions $g^{-1}_{\nu,D}$ forms a convex set. However, Eq.~ (\ref{eq:inverwarpingmap:convecdiff}) indicates that the set $\mathcal{G}_{\nu,D}^{-1}$ is non-convex with respect to $D$. Nonetheless, it can be shown that within a small neighborhood of $D_0$ (that is for $D \in [D_0-\epsilon,D_0+\epsilon]$), the set of solutions is approximately convex in SCDT space (see \ref{app:conv-diff_convex}). 

Similarly, the family of inverse warping maps defined in Eq.~(\ref{eq:inverwarpingmap:convecdiff}) can also be shown to approximately form a convex set about the $\nu$ parameter (see \ref{app:conv-diff_convex}). 
Hence, the transport-based modeling method can be implemented to recover either of the two parameters in the convection-diffusion equation, with the other parameter allowed to vary within a small range. Next, we will delve into how this technique can address parameter identification problems for situations where deriving an analytical solution to the system PDE model is challenging.

\section{Proposed Method} \label{sec:method}
The aim of this paper is to demonstrate a practical and efficient solution for system identification problems by using a transport transform-based modeling approach. As demonstrated earlier, these types of mathematical models can be derived analytically for various wave propagation phenomena, such as the wave equation, diffusion equation, convection-diffusion equation, etc. Nevertheless, for situations where deriving an analytical model is challenging, we adopt a data-driven approach to learn the signal classes. This section outlines our proposed data-driven approach for addressing system identification problems.

\subsection{Data-driven Identification of PDE Parameters}
Consider a dynamical system that describes 1D propagation of waves through a potentially damaged, elastic medium. A PDE model for this system can be expressed as follows:
\begin{equation}\label{eq:we_damaged}
    \rho \ddot{u}-Eu_{xx}-\eta \dot{u}_{xx}-\rho M \ddot{u}_{xx} - Fu_{xxxx} + \mathcal{B} u_x u_{xx} = \Phi,
\end{equation}
where the first two terms on the left-hand side describe classical, d'Alembertian wave propagation, the remaining terms on the left-hand side incorporate several mechanisms that distort traveling waves, and the right-hand side models an externally applied excitation. For the present study, we interpret and discuss Eq.~\eqref{eq:we_damaged} as a model for longitudinal wave propagation in a thin, elastic rod \cite{graff2012wave}. In this context\footnote{Note that our chosen mathematical symbols generally differ from those used in the cited references.}, $ u $ is the displacement parallel to the direction of propagation, $ \rho $ is the mass density, $ E $ is a general, elastic modulus, which may be interpreted as Young's modulus in the case of a homogeneous rod \cite{graff2012wave}, and $ \Phi $ is a force per unit volume. The terms proportional to $ \eta $ and $ \mathcal{B} $ incorporate linear damping (using a Kelvin-Voigt viscoelastic model; see section 7.18 in \cite{meirovitch1997}) and quadratic, elastic nonlinearity \cite{norris1998finite}, respectively. Finally, the terms proportional to $ M $ and $ F $ include short-wavelength dispersion that arises due to subwavelength, geometrical effects, such as lateral inertia  and periodic microstructure (see, e.g., section 2.5 in \cite{graff2012wave}, chapter 1 in \cite{ablowitz2011nonlinear}, \cite{metrikine2002}, and references therein).

For a given physical system, the numerical values of the coefficients in Eq.~\eqref{eq:we_damaged} depend on the materials and the characteristic length and time scales present, and may generally span many orders of magnitude; this presents a potential challenge for data-driven approaches, as a set of training data may be applicable only to a narrow range of possible parameters. However, by re-scaling the variables $ u $, $ x $, and $ t $, Eq.~\eqref{eq:we_damaged} may be re-written in the following dimensionless form:
\begin{equation}
    \xi_0 \ddot{u}-\xi_1 u_{xx}-\xi_2 \dot{u}_{xx}-\xi_3 \ddot{u}_{xx} - \xi_4 u_{xxxx}+\beta u_x u_{xx} = \phi,
    \label{eq:we_gen}
\end{equation}
where $\boldsymbol{\xi} = [\xi_0, \xi_1, \xi_2, \xi_3, \xi_4]^T$ are the coefficients of the linear terms of the PDE and are each $ O(1) $, $ \beta $ is the dimensionless coefficient of nonlinearity, and $ \phi $ is the dimensionless force. Thus, one could potentially generate a training set by varying all the PDE coefficients through a relatively narrow range, and apply it to experimental signals from a broad variety of physical systems by scaling the data appropriately.	The nondimensionalization procedure is detailed in \ref{app:nondim}. 

In numerous structural health monitoring (SHM) applications, the process of system identification involves the retrieval of the nonlinearity parameter ($\beta$), as it is frequently linked to the existing damage in the structure \cite{van2000nonlinear,Farrar07,Nichols:07}. Hence, the primary emphasis in this section lies on parameter $\beta$, which we aim to recover by employing sensor data $s(t)$ measured at a particular location $x=x_m$.

\subsection{Signal Class Model and Problem Statement}
This section aims to propose a mathematical model-based problem statement to identify the system parameter $\beta$ using the sensor measurement $s(t)$. As stated in equation \ref{eqn:transportJMN}, the assumption is that each measured signal can be modeled as an instance of a particular template observed under some unknown deformations. Such a family of signals can be described with the following model:

\subsubsection{Signal Class Model}
Let the coefficients in $\boldsymbol{\xi}$ vary within a given range, such that $\xi_i\in[\mathcal{E}_i-\epsilon, \mathcal{E}_i+\epsilon]$ for $i=0,1,...,4$ and $\mathcal{E}_i\in \mathbb{R}$. Let $\mathcal{G}^{(\beta)}_{\boldsymbol{\xi}} \subset \mathcal{T}$ denote a set of increasing 1D deformations of a specific kind, where $\mathcal{T}:\mathbb{R} \rightarrow\mathbb{R}$ is a set of all possible increasing diffeomorphisms. Given a $\beta$, there exists a template pattern $\varphi^{(\beta)}_{\boldsymbol{\xi}}(x,t)$ and a warping function $g_j(x,t)\in \mathcal{G}^{(\beta)}_{\boldsymbol{\xi}}$ for a small $\epsilon$ such that the family of the sensor measurement $s(t)$ at a fixed location $x=x_m$ can be modeled as:
\begin{equation}
    \mathbb{S}_{\varphi^{(\beta)}_{\boldsymbol{\xi}},\mathcal{G}^{(\beta)}_{\boldsymbol{\xi}}} = \{s_j^{(\beta)}|s_j^{(\beta)}=\dot{g}_j\varphi^{(\beta)}_{\boldsymbol{\xi}}\circ g_j, g_j\in \mathcal{G}^{(\beta)}_{\boldsymbol{\xi}},\dot{g}_j>0\},
    \label{eq:gen_mod0}
\end{equation}
where $s^{(\beta)}_j$ is the $j$-th signal under a given nonlinearity ($\beta$), and $\dot{g}_j$ represents first order derivative of $g_j$ with respect to $t$. It is also assumed that for a small $\epsilon$, the family of functions $g^{-1}_j\in \left(\mathcal{G}^{(\beta)}_{\boldsymbol{\xi}}\right )^{-1},\,\forall j$ forms a convex set.
Note that since the sensor location $x$ is fixed at $x_m$, the signals $s_j$, $\varphi^{(\beta)}_{\boldsymbol{\xi}}$, and the warping functions $g_j$ are represented as a function of time $t$ only, e.g., $\varphi^{(\beta)}_{\boldsymbol{\xi}}(t)=\varphi^{(\beta)}_{\boldsymbol{\xi}}(x=x_m,t)$, $g_j(t)=g_j(x=x_m,t)$, etc.

The warping function $g_j(\cdot)$ in the signal class model defined by Eq.~(\ref{eq:gen_mod0}) transforms the independent variable (time, $t$ in this problem) according to the structural dynamics (including damage), and the effect of this function is to morph the template signal $\varphi^{(\beta)}_{\boldsymbol{\xi}}$ into the observed signal $s_j^{(\beta)}$. 
Many wave propagation phenomena, accurately represented by the wave equation, diffusion equation, and convection-diffusion equation, can be described using such models. However, for PDEs like the one described in Eq.~(\ref{eq:we_damaged}), deriving an analytical solution is challenging, making it difficult to analytically find a single template for the signal class. In this paper, we assume that we can find a template locally and hence, the sensor measurements from a single class can be modeled using the multiple template-based signal class model defined in Eq.~(\ref{eq:genmod_multi}).

\subsubsection{Problem Statement}\label{sec:prob_st}
Considering the signal class model stated above, the system parameter identification problem for the dynamical system described by the PDE shown in Eq.~(\ref{eq:we_gen}) is defined as follows:

\noindent\textbf{Identification problem:} \textit{Let $\mathcal{G}^{(\beta)}_{\boldsymbol{\xi}} \subset \mathcal{T}$ be a set of increasing temporal deformations, and $\mathbb{S}_{\varphi^{(\beta)}_{\boldsymbol{\xi}},\mathcal{G}^{(\beta)}_{\boldsymbol{\xi}}}$ be defined as in Eq.~(\ref{eq:gen_mod0}). Given a set of sensor measurements $\{s_1^{(\beta)}, s_2^{(\beta)},...\}\subset \mathbb{S}_{\varphi^{(\beta)}_{\boldsymbol{\xi}},\mathcal{G}^{(\beta)}_{\boldsymbol{\xi}}}$ with known $\beta\in\mathbb{R}$, identify the system parameter $\beta$ for an unknown measurement $s$. }

In this paper,  we further split this identification problem into two sub-problems:
\begin{itemize}
    \item \textbf{Detection of the parameter:} The detection problem for the parameter $\beta$ can be framed as a binary classification problem where classes 1 and 2 represent the sensor data corresponding to $\beta=0$ and $\beta\neq 0$, respectively. Signals from class $c$ ($c=1,2$) can be modeled as:
    \begin{align} 
        &\mathbb{S}^{(c)} = \bigcup\limits_{m=1}^{M_c} \mathbb{S}_{\varphi^{(c)}_{m},\mathcal{G}^{(c)}_{m}}, \nonumber \\
        &\mathbb{S}_{\varphi^{(c)}_{m},\mathcal{G}^{(c)}_{m}} = \{s_{j,m}^{(c)}|s_{j,m}^{(c)}=\dot{g}_j\varphi^{(c)}_{m}\circ g_j, g_j\in \mathcal{G}^{(c)}_{m},\dot{g}_j>0\},
        \label{eq:gen_mod1}
    \end{align}
    where $\varphi^{(c)}_{m}$ is the $m$-th template pattern from class $c$, and $s_{j,m}^{(c)}$ is the $j$-th signal generated from $\varphi^{(c)}_{m}$ under the deformation defined by $g_j$. 
    Given the signal class model, the detection problem can be defined as follows:
    
    \textit{Let $\mathcal{G}^{(c)}_{m} \subset \mathcal{T}$ be a set of increasing temporal deformations, and $\mathbb{S}^{(c)}$ be defined as in Eq.~(\ref{eq:gen_mod1}), for classes $c=1,2$. Given a set of sensor measurements $\{s_1^{(c)}, s_2^{(c)},...\}\subset \mathbb{S}^{(c)}$ for class $c$, determine the class label of an unknown signal $s$, meaning, detect if the parameter of interest ($\beta$) is zero or not.}
    
    \item \textbf{Estimate the parameter value:} The objective is to estimate the value of the system parameter $\beta$ for an unknown measurement $s$, given a set of sensor measurements $\{s_1^{(\beta)}, s_2^{(\beta)},...\}\subset \mathbb{S}_{\varphi^{(\beta)}_{\boldsymbol{\xi}},\mathcal{G}^{(\beta)}_{\boldsymbol{\xi}}}$ with known $\beta$ values. However, obtaining a precise estimation of $\beta$ requires a large number of training samples from the set $\mathbb{S}_{\varphi^{(\beta)}_{\boldsymbol{\xi}},\mathcal{G}^{(\beta)}_{\boldsymbol{\xi}}}$, which may not be feasible in a real-world scenario. In this work, the estimation of the nonlinearity parameter is framed as a coarse regression problem where a range $[l_{\beta}, h_{\beta}]$ is predicted for the unknown $\beta$. 
    Furthermore, this regression problem can be viewed as a multi-class classification problem, where signals from class $c$ (for $c=1,2,..., N_c$) can be modeled by the model defined in Eq.~(\ref{eq:gen_mod1}) and the set $\mathbb{S}_{\varphi^{(c)}_{m},\mathcal{G}^{(c)}_{m}}$ corresponds to $\beta_m^{(c)}\in[l_{\beta_m^{(c)}}, h_{\beta_m^{(c)}}]$. Given the signal class model, the problem of estimating the system parameter $\beta$ is defined as follows:
    
    \textit{Let $\mathcal{G}^{(c)}_{m} \subset \mathcal{T}$ be a set of increasing temporal deformations, and $\mathbb{S}^{(c)}$ be defined as in Eq.~(\ref{eq:gen_mod1}), for classes $c=1, 2, ..., N_c$. Given a set of sensor measurements $\{s_1^{(c)}, s_2^{(c)},...\}$ $\subset \mathbb{S}^{(c)}$ for class $c$, determine the class label of an unknown signal $s$, meaning, estimate a range $[l_{\beta}, h_{\beta}]$  for the system parameter $\beta$.}
\end{itemize}

\subsection{Proposed Solution} \label{sec:solution}
The proposed solution to the regression problems defined earlier involves using the SCDT-NLS method \cite{rubaiyat2022end} described in section \ref{sec:scdt-nls}. The signal class model defined in Eq.~(\ref{eq:gen_mod1}) typically results in nonconvex signal classes, making the classification problems challenging to solve. However, as explained in section \ref{sec:SCDT}, we can simplify the geometry of the signal classes under certain assumptions by using the SCDT. Hence, the proposed solution begins with applying the SCDT on the sensor measurements. The signal class model in SCDT domain is:
\begin{align}
    &\widehat{\mathbb{S}}^{(c)} = \bigcup\limits_{m=1}^{M_c} \widehat{\mathbb{S}}_{\varphi^{(c)}_{m},\mathcal{G}^{(c)}_{m}}, \nonumber\\
    &\widehat{\mathbb{S}}_{\varphi^{(c)}_{m},\mathcal{G}^{(c)}_{m}} = \{\widehat{s}_{j,m}^{(c)}|\widehat{s}_{j,m}^{(c)}=g^{-1}_j\circ\widehat{\varphi}^{(c)}_{m}, g^{-1}_j\in \left(\mathcal{G}^{(c)}_{m}\right)^{-1}\},
    \label{eq:gen_mod1_scdt}
\end{align}
where $g^{-1}_j\circ\widehat{\varphi}^{(c)}_{m}$ refers to the SCDT of the signal $\dot{g}_j\varphi^{(c)}_{m}\circ g_j$. In many wave propagation phenomena, such as those described by the wave equation, convection-diffusion equation, etc., the set $\left(\mathcal{G}^{(c)}_{m}\right)^{-1}$ can be shown (or approximated) to be convex. As a result, by utilizing the convexity property of the SCDT outlined in section \ref{sec:SCDT}, we can demonstrate that the set $\widehat{\mathbb{S}}_{{\varphi^{(c)}_{m},\mathcal{G}^{(c)}_{m}}}$ defined in equation (\ref{eq:gen_mod1_scdt}) forms a convex set. Moreover, since the SCDT is a one-to-one mapping, if $\mathbb{S}_{\varphi_m^{(c)},\mathcal{G}_m^{(c)}} \cap \mathbb{S}_{\varphi_w^{(p)},\mathcal{G}_w^{(p)}}=\varnothing$ for $c\neq p$, then $\widehat{\mathbb{S}}_{\varphi_m^{(c)},\mathcal{G}_m^{(c)}} \cap  \widehat{\mathbb{S}}_{\varphi_w^{(p)},\mathcal{G}_w^{(p)}}=\varnothing$.

As outlined in section \ref{sec:scdt-nls}, the class label of an unknown test sample $s$ can be predicted by solving,
\begin{equation}
    \argmin_c\min_m~d^2\left(\widehat{s}, \widehat{\mathbb{V}}_m^{(c)}\right), 
    \label{eq:min_problem}
\end{equation}
where $\widehat{\mathbb{V}}_m^{(c)} = \text{span}\left(\widehat{\mathbb{S}}_{\varphi_m^{(c)},\mathcal{G}_m^{(c)}}\right)$.
We then exploit the nearest local subspace search algorithm (outlined in \cite{rubaiyat2022end}) in SCDT domain to solve this classification problem.

Consider a set of training samples $\left\{s_1^{(c)},..., s_j^{(c)},...,s_{L_c}^{(c)}\right\}\subset \mathbb{S}^{(c)}$ for class $c$, where $L_c$ is the total number of training samples given for class $c$, and $s_j^{(c)}$ is the $j$-th sample. The training and testing phases of the algorithm are described below:

\subsubsection{Training Phase}
Training phase begins with computing the SCDTs for all training samples from class $c$. Then, we take a training sample $\widehat{s}_{l}^{(c)}$ and obtain the basis vector that spans the subspace corresponding to that sample.
Let $B_{l}^{(c)} = \left[b_{l}^{(c)} \right]$ be a matrix that contains the basis vector in its column. These calculations are repeated for all the training samples to form $B_{l}^{(c)}$ for $l=1,2,...,L_c$ and $c=1,2,...,$ etc.

\subsubsection{Testing Phase}
The testing phase begins with taking SCDT of the test sample $s$ to obtain $\widehat{s}$ followed by the nearest local subspace search in SCDT domain. In the first step of the algorithm, we estimate the distance of the subspace corresponding to each of the training samples from $\widehat{s}$ by:
\begin{align*}
    \epsilon_l = \|\widehat{s} - B_{l}^{(c)}B_{l}^{(c)^T}\widehat{s}\|^2,\qquad l=1,2,...,L_c,
\end{align*}
where $\|\cdot\|$ denotes $\ell_2$ norm.
Subsequently, a set of $k$ training samples closest to the test sample $\widehat{s}$ from class $c$, denoted by $\{\widehat{s}_1^{(c)},\cdots,\widehat{s}_k^{(c)}\}$, is found based on the distances $\epsilon_1, \epsilon_{2}, ...,$ etc.
Next, $\{\widehat{s}_1^{(c)},\cdots,\widehat{s}_k^{(c)}\}$ is orthogonalized to obtain the basis vectors $\{b_1^{(c)},b_2^{(c)},...\}$ spanning the local subspace from class $c$ with respect to $\widehat{s}$. The unknown class of $s$ is then estimated by:
\begin{equation}
    \arg\min_c~ \|\widehat{s} - B^{(c)}B^{(c)^T}\widehat{s}\|^2,
    \label{eq:test_step}
\end{equation}
where $B^{(c)}=\left[b_1^{(c)},b_2^{(c)},... \right]$. The presence and extent of damage-induced nonlinearity within the propagation medium is determined by the predicted class. 

In this section, we outlined a mathematical modeling-based data-driven approach to recover the nonlinearity parameter $\beta$ of the dynamical system defined by Eq.~(\ref{eq:we_damaged}). This approach can also be extended to identify other system parameters such as dispersion parameters $M$ or $F$, as well as the dissipation parameter $\eta$ and so on.
It is important to note that the proposed solution does not require knowledge of the template patterns or the warping functions present in the data. The approach utilizes a set of training samples to search for the nearest local subspace for a given test sample. In the case of dynamical systems where the PDEs are known, we can numerically simulate the PDE solutions and use the simulated data during the training phase. If the governing PDE is unknown, previously acquired data from a controlled experimental setup can be utilized as training samples.

\section{Numerical Experiments and Results} \label{sec:experiments}

\subsection{Experimental Setup}\label{sec:setup}
\begin{figure}[tb]
    \centering
    \includegraphics[width=0.85\textwidth]{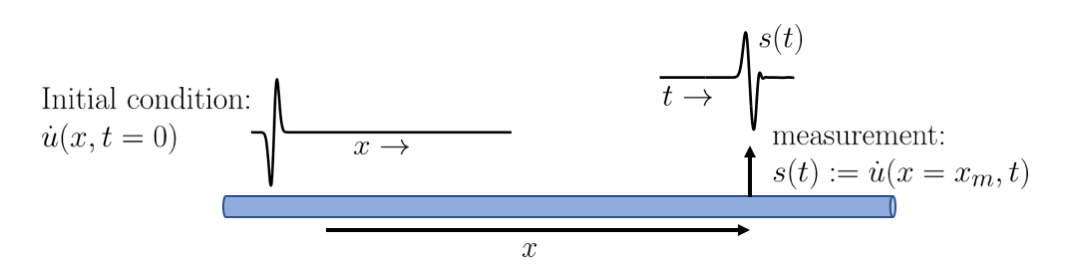}
    \caption{Simulation setup for 1D wave propagation through an elastic medium.}
    \label{fig:1d_dyn_sys_exp}
    \vspace{-1.0em}
\end{figure}

Since obtaining an analytical solution to the PDE defined in Eq.~(\ref{eq:we_damaged}) is difficult, we resorted to numerically simulating the PDE solution using the spectral method \cite{kopriva2009implementing}. For this purpose, we utilized a fast Fourier transform-based implementation \cite{brunton2022data} of the PDE simulator, which mandates the use of periodic boundary conditions. Moreover, we defined a set of initial conditions instead of using a source function to initiate the wave propagation. The initial conditions for the displacement and the velocity are given by:
\begin{align}
    &u(x,0) = e^{-\frac{(x-x_0)^2}{2\sigma^2}} \nonumber\\
    &\dot{u}(x,0) = \frac{\nu(x - x_0)}{\sigma^2}u(x,0),
    \label{eq:init_cond}
\end{align}
where $\dot{u}(x,0):=\frac{\partial u(x,t)}{\partial t}|_{t=0}$ and $\nu$ is the wave speed. The selection of the initial velocity $\dot{u}(x,0)$ was made to produce a pure traveling wave in $+x$ direction for the standard wave equation. A spatial grid of $600$ points was used, mean location $x_0$ and standard deviation $\sigma$ were set to $50$ and $7$, separately. The wave speed, which is given by $\nu = \sqrt{\frac{E}{\rho}}$, was computed with the mass density of the medium $\rho$ set at 1, and Young's modulus $E$ chosen from a uniform distribution $U(0.95,1.05)$. In addition, the parameter $F$ was fixed at 0.01, while $M$ and $\eta$ were both selected from the distributions $U(0.2,0.3)$ and $U(0.1,0.2)$, respectively. In this particular experiment, the velocity $\dot{u}(x,t)$ was considered as the primary quantity of interest and was measured as a function of time at a fixed sensor location, $x=300$, i.e., the sensor measurement $s(t)=\dot{u}(x=300,t)$. The simulation setup for this experiment is depicted in Fig.~\ref{fig:1d_dyn_sys_exp}.

\begin{figure}[tb]
    \centering
    \includegraphics[width=0.8\textwidth]{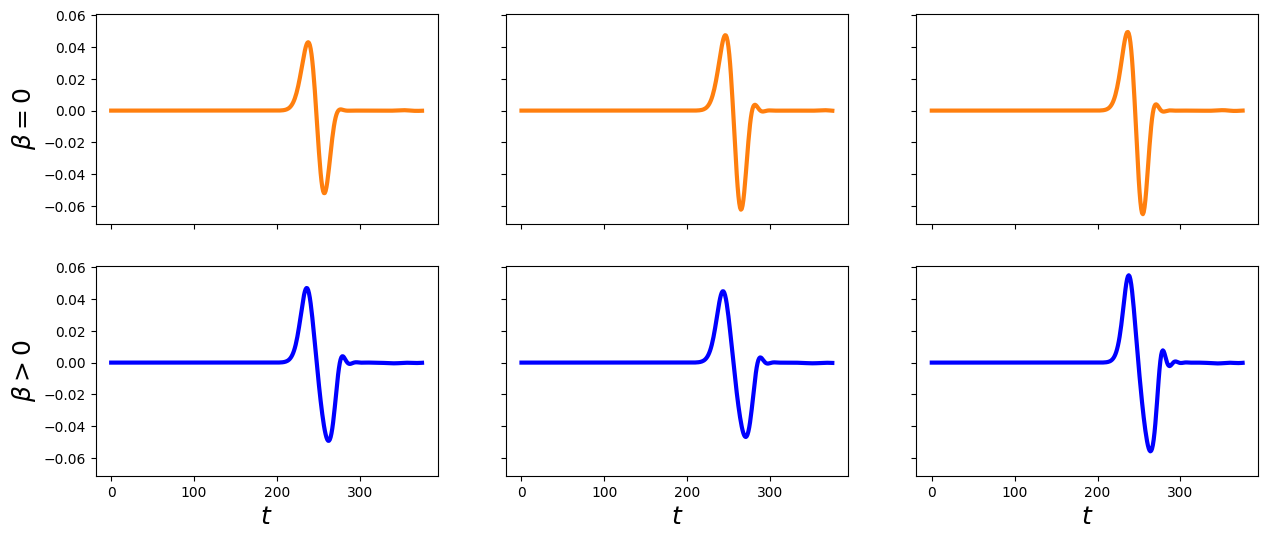}
    \caption{Sensor measurements at a particular location (top row) without nonlinearity, and (bottom row) with nonlinearity.}
    \label{fig:sig_2class}
    \vspace{-1.0em}
\end{figure}

\begin{figure}[tb]
    \centering
    \includegraphics[width=0.8\textwidth]{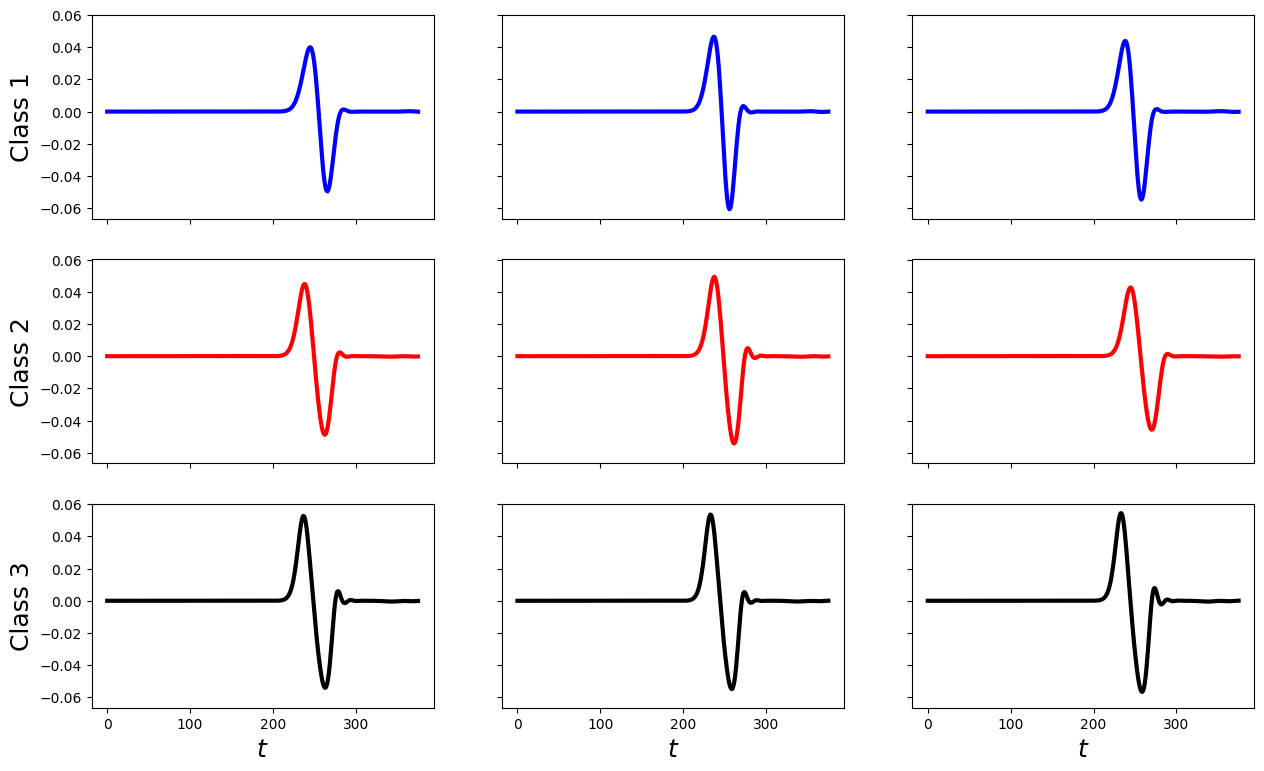}
    \caption{Sensor measurements at a particular location under the presence of three different levels of nonlinearity.}
    \label{fig:sig_3class}
    \vspace{-1.0em}
\end{figure}

The nonlinearity detection problem was set up as a binary classification problem, where class 1 denotes signals corresponding to $\beta=0$ (no nonlinearity) and class 2 consists of signals produced by randomly changing $\beta$ (randomly chosen from a uniform distribution $U(0.01,0.6)$). Fig.~\ref{fig:sig_2class} displays a number of sensor measurements, characterized by the absence ($\beta = 0$) or presence ($\beta > 0$) of nonlinearity.
To assess the degree of nonlinearity, a coarse regression problem was formulated, in which the proposed method predicts a range of $\beta$ values for an unknown signal. In the initial experiment, the sensor measurements were grouped into three categories based on their corresponding $\beta$ values (randomly chosen from $U(0.01,0.2)$, $U(0.21,0.4)$, and $U(0.41,0.6)$, respectively). Fig.~\ref{fig:sig_3class} presents a set of example signals featuring three distinct levels of nonlinearity. Subsequently, this experiment was converted into a finer regression setup with ten classes, where a narrower range of $\beta$ values was predicted for an unknown sensor measurement. Table \ref{table:10class_beta} displays the ten distributions that correspond to the ten classes. The classification method described in section \ref{sec:method} was then utilized to identify the nonlinearity parameter $\beta$ from the sensor measurements.

\begin{table}[!htb]
    \centering
    \caption{Coarse regression (10-class) problem setup for estimating $\beta$. Here, $\beta$ is randomly chosen from the uniform distribution $U(l_\beta,h_\beta)$, where $l_\beta$ and $h_\beta$ corresponding to ten different classes are shown in this table.}
    {\small
    \begin{tabular}{lcccccccccc}
    \hline
        Class & 1 & 2 & 3 & 4 & 5 & 6 & 7 & 8 & 9 & 10 \\
        \hline
         $l_\beta$ & $0.01$ & $0.07$ & $0.13$ & $0.19$ & $0.25$ & $0.31$ & $0.37$ & $0.43$ & $0.49$ & $0.55$ \\
        \hline
        $h_\beta$ & $0.06$ & $0.12$ & $0.18$ & $0.24$ & $0.30$ & $0.36$ & $0.42$ & $0.48$ & $0.54$ & $0.6$ \\
        \hline
    \end{tabular}
    }
    \label{table:10class_beta}
\end{table}

\subsection{Identification of nonlinearity parameter}
We conducted a comparative analysis of the proposed method with linear-support vector machine (SVM) \cite{hearst1998support} classifier and some state-of-the-art deep neural network techniques, including Multilayer Perceptrons (MLP) \cite{iwana2021empirical}, 1D Visual Geometry Group (VGG) \cite{iwana2021empirical}, 1D Residual Network (ResNet) \cite{fawaz2018data} \cite{iwana2021empirical}, 
and Long Short Term Memory Fully Convolutional Network (LSTM-FCN) \cite{karim2017lstm} \cite{iwana2021empirical}. We also implemented a Fourier transform-based method as a traditional approach, which employs a linear-SVM classifier to classify the sensor measurements in the Fourier domain. Moreover, we compared the performance against SCDT-NS \cite{rubaiyat2022nearest}, which models each signal class as a single subspace in the SCDT domain.

\subsubsection{Detecting nonlinearity}
The PDE simulation procedure described in section \ref{sec:setup} was executed 2200 times per class with randomly varying parameters. The sensor signal was measured at a single location during the simulation, resulting in 2200 sensor measurements per class. The collected data were then split into two sets, with 2000 signals per class for training and 200 signals per class for testing. The classification models listed above were then trained and evaluated. The performance of detecting nonlinearity is reported as classification accuracy in Table \ref{table:nl_detect_acc}. The table shows that the proposed solution achieved a nonlinearity detection accuracy of $98.0\%$, which is higher than that of the comparative methods. These results suggest that the proposed approach has the potential to detect nonlinearity induced by damage in the propagation medium.

\begin{table}[tb]
    \centering
    \caption{Nonlinearity detection accuracy ($\%$) for different classification methods. All the models were trained using 2000 samples per class, and tested on 200 samples per class.}
    {\small
    \begin{tabular}{|l|c|}
    \hline
        Methods & Accuracy (\%) \\
        \hline
        MLP & 73.8 \\
        1D-VGG & 91.9 \\
        1D-ResNet & 63.2 \\
        LSTM-FCN & 83.6 \\
        SVM & 79.8 \\
        FT-SVM & 95.5 \\
        SCDT-NS & 91.5 \\
        \textbf{SCDT-NLS} & \textbf{98.0} \\
        \hline
    \end{tabular}
    }
    \label{table:nl_detect_acc}
\end{table}

\begin{figure}[tb]
    \centering
    \includegraphics[width=0.75\textwidth]{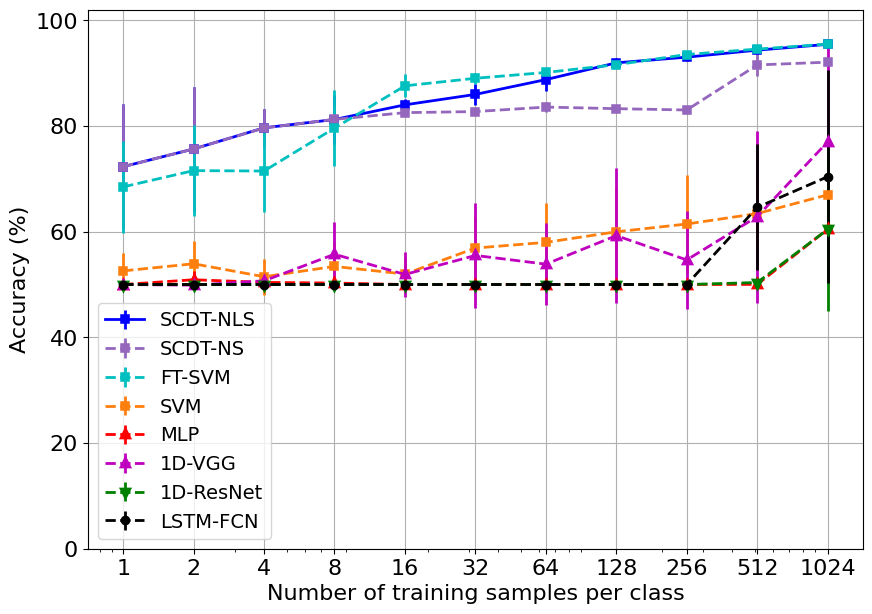}
    \caption{Nonlinearity detection accuracy as a function of number of training samples per class for different classification methods.}
    \label{fig:det_beta}
    \vspace{-1.0em}
\end{figure}

In addition to its effectiveness, the proposed solution is also data-efficient, allowing it to achieve superior detection accuracy using fewer training samples than other methods. In order to demonstrate the data efficiency of the proposed approach, an experiment was conducted where all classifiers were trained with varying numbers of training samples per class. A training split of a certain size was randomly selected from the original training set, and the experiments for this particular size were repeated 10 times. The results are presented in Fig.~\ref{fig:det_beta}, which displays the average detection accuracy as a function of the number of training samples per class for the different classification methods. The error bars indicate the standard deviation for each split. These plots demonstrate that the proposed method outperforms the comparative methods in terms of accuracy, even with a smaller number of training samples.

\subsubsection{Estimating the degree of nonlinearity}
\begin{figure}[tb]
    \centering
    \includegraphics[width=0.75\textwidth]{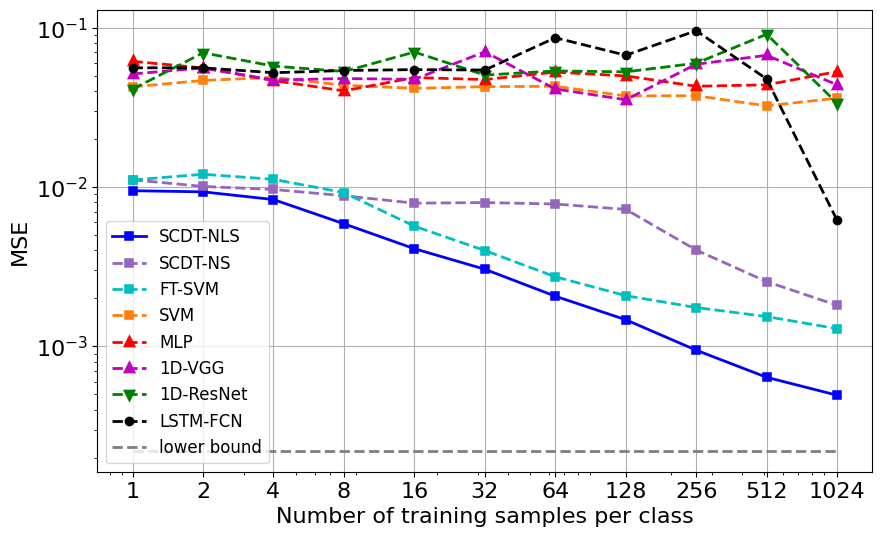}
    \caption{Nonlinearity estimation performance (MSE) as a function of number of training samples per class for different methods under coarse regression (10-class) setup. Plot labeled as `lower bound' represents the least possible MSE for this experiment.}
    \label{fig:est_beta10}
    \vspace{-1.0em}
\end{figure}

\begin{table}[tb]
    \centering
    \caption{Nonlinearity estimation error in MSE (Eq.~\ref{eq:est_mse}) for different methods. All the models were trained using 2000 samples per class, and tested on 200 samples per class.}
    {\small
    \begin{tabular}{|l|c|c|}
    \hline
        Methods & MSE (3-class) & MSE (10-class) \\
        \hline
        MLP & $1.87\times 10^{-2}$ & $3.12\times 10^{-2}$ \\
        1D-VGG & $5.10\times 10^{-3}$ & $1.71\times 10^{-2}$ \\
        1D-ResNet & $3.72\times 10^{-2}$ & $3.84\times 10^{-2}$ \\
        LSTM-FCN & $1.25\times 10^{-2}$ & $9.77\times 10^{-3}$ \\
        SVM & $1.99\times 10^{-2}$ & $2.67\times 10^{-2}$ \\
        FT-SVM & $3.66\times 10^{-3}$ & $9.84\times 10^{-4}$ \\
        SCDT-NS & $4.30\times 10^{-3}$ & $1.41\times 10^{-3}$ \\
        \textbf{SCDT-NLS} & $\boldsymbol{3.14\times 10^{-3}}$ & $\boldsymbol{3.08\times 10^{-4}}$ \\
        \hline
    \end{tabular}
    }
    \label{table:nl_est_mse}
\end{table}
As described in section \ref{sec:prob_st}, the estimation of nonlinearity parameter $\beta$ is stated as a classification problem, where the proposed approach predicts a range for $\beta$. Initially, a three-class classification problem was set up based on the values of $\beta$. Afterwards, a finer regression problem was created, where the sensor measurements were classified into ten different classes, and $\beta$-values were selected from ten non-overlapping distributions as shown in Table \ref{table:10class_beta}. In both scenarios, the classification models were trained using 2000 samples per class, collected by simulating the PDE, and then used to predict $\beta$-ranges for 200 test samples per class. To evaluate the performance of the methods, the mean squared errors (MSE) were calculated as follows:
\begin{equation}
    \mathrm{MSE} = \frac{1}{N}\sum_{i=1}^N \left(\beta_i - \frac{l_{\beta_i}+h_{\beta_i}}{2}\right)^2,
    \label{eq:est_mse}
\end{equation}
where $N$ represents the total number of test samples, $\beta_i$ is the true nonlinearity parameter value of the $i$-th test sample, and $l_{\beta_i}$ and $h_{\beta_i}$ are the lower and upper limits of the predicted range, respectively. The performance of the classification methods for estimating $\beta$ under both regression problem setups is presented in Table \ref{table:nl_est_mse}. According to the results, the proposed approach yields the least MSE in both cases, suggesting that it outperforms all other methods in accurately estimating the degree of nonlinearity present in the propagation medium. Fig. \ref{fig:est_beta10} illustrates the nonlinearity estimation performances of different methods as the number of training samples varies. The plots in the figure highlight that SCDT-NLS outperforms the other methods in estimating the degree of nonlinearity with fewer number of  training samples.

\subsection{Identification of dispersion parameter}
We utilized the proposed system identification technique to recover the dispersion parameter $M$ for the dynamical system defined by Eq.~(\ref{eq:we_damaged}). Following a similar approach as with the parameter $\beta$, we formulated a binary classification problem: class 1 represents signals associated with $M=0$, while class 2 consists of signals corresponding to randomly selected $M$ values from the uniform distribution $U(0.01,0.6)$. We then employed the regression methods listed above to detect the dispersion parameter $M$. The results in Table \ref{table:M_identify} (second column) demonstrate that the proposed approach outperforms state-of-the-art regression methods in detecting the dispersive dynamical systems. To assess the degree of dispersion, a coarse regression problem was formulated where the sensor measurements were categorized into three groups based on the values of $M$ randomly chosen from distributions $U(0.01,0.2)$, $U(0.21,0.4)$, and $U(0.41,0.6)$. The comparative performance of the proposed approach with respect to the existing methods is shown in the third column of Table \ref{table:M_identify}. The results suggest that the proposed system identification approach outperforms all other regression methods in accurately estimating the degree of dispersion.

\begin{table}[tb]
    \centering
    \caption{Identification of dispersion parameter: 2nd column shows the accuracy (in $\%$) of detecting the presence of dispersion (2-class problem) for different methods. 3rd column shows the dispersion parameter estimation error in MSE (3-class problem) for different methods. All the models were trained using 2000 samples per class, and tested on 200 samples per class.}
    {\small
    \begin{tabular}{|l|c|c|}
    \hline
        Methods & Detection Accuracy (\%) & Estimation Error (MSE) \\
        \hline
        MLP & 63.3 & $3.17\times 10^{-2}$ \\
        1D-VGG & 55.3 & $4.26\times 10^{-2}$ \\
        1D-ResNet & 50.2 & $6.28\times 10^{-2}$ \\
        LSTM-FCN & 74.5 & $8.65\times 10^{-3}$ \\
        SVM & 63.25 & $5.08\times 10^{-2}$ \\
        FT-SVM & 69.75 & $1.28\times 10^{-2}$ \\
        SCDT-NS & 96.75 & $8.1\times 10^{-3}$ \\
        \textbf{SCDT-NLS} & \textbf{99.0} & $\boldsymbol{3.13\times 10^{-3}}$ \\
        \hline
    \end{tabular}
    }
    \label{table:M_identify}
\end{table}

\section{Application: Structural Health Monitoring} \label{sec:data_experiment}
\begin{figure}[tb]
    \centering
    \includegraphics[width=0.8\textwidth]{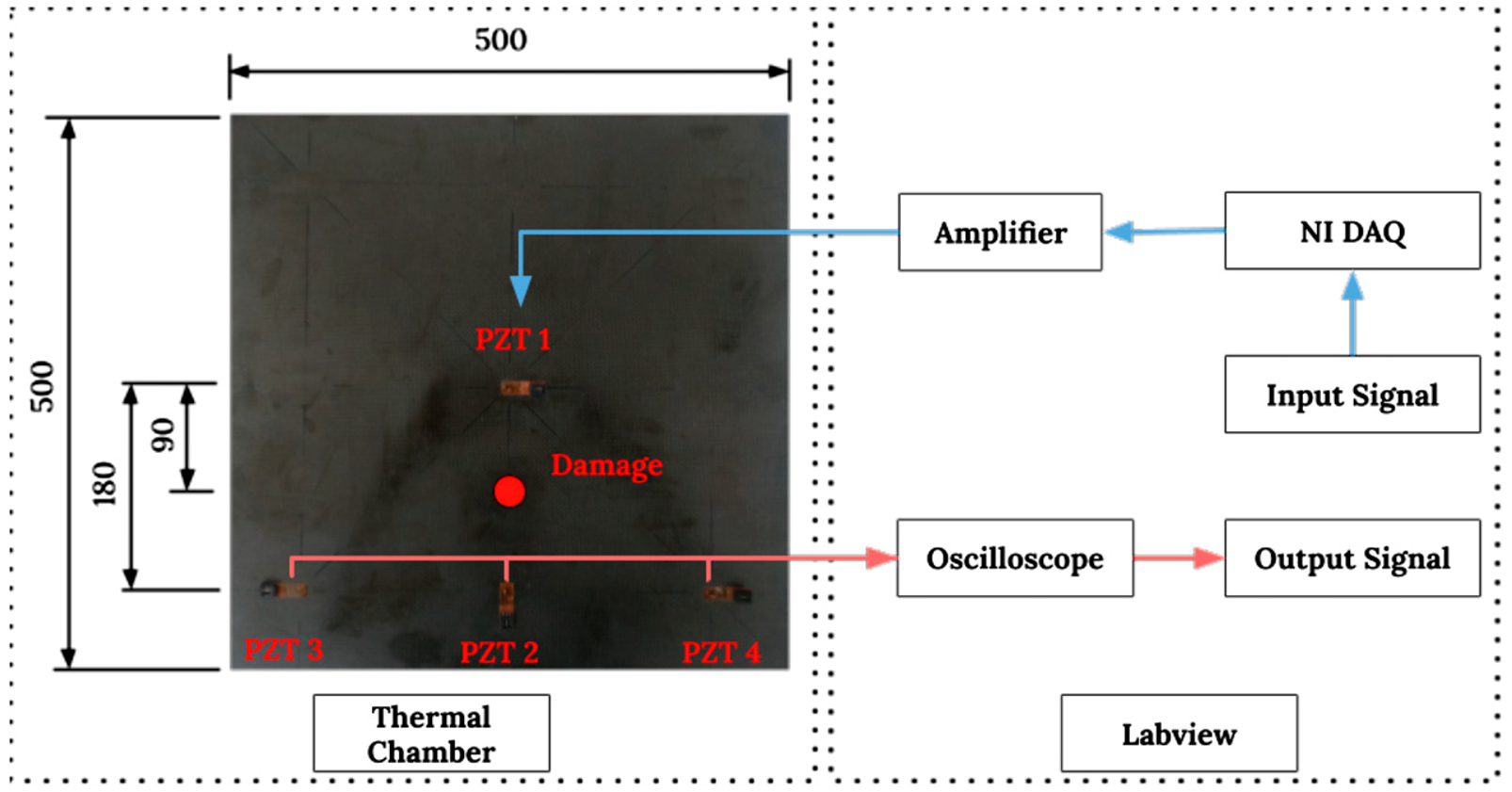}
    \includegraphics[width=0.6\textwidth]{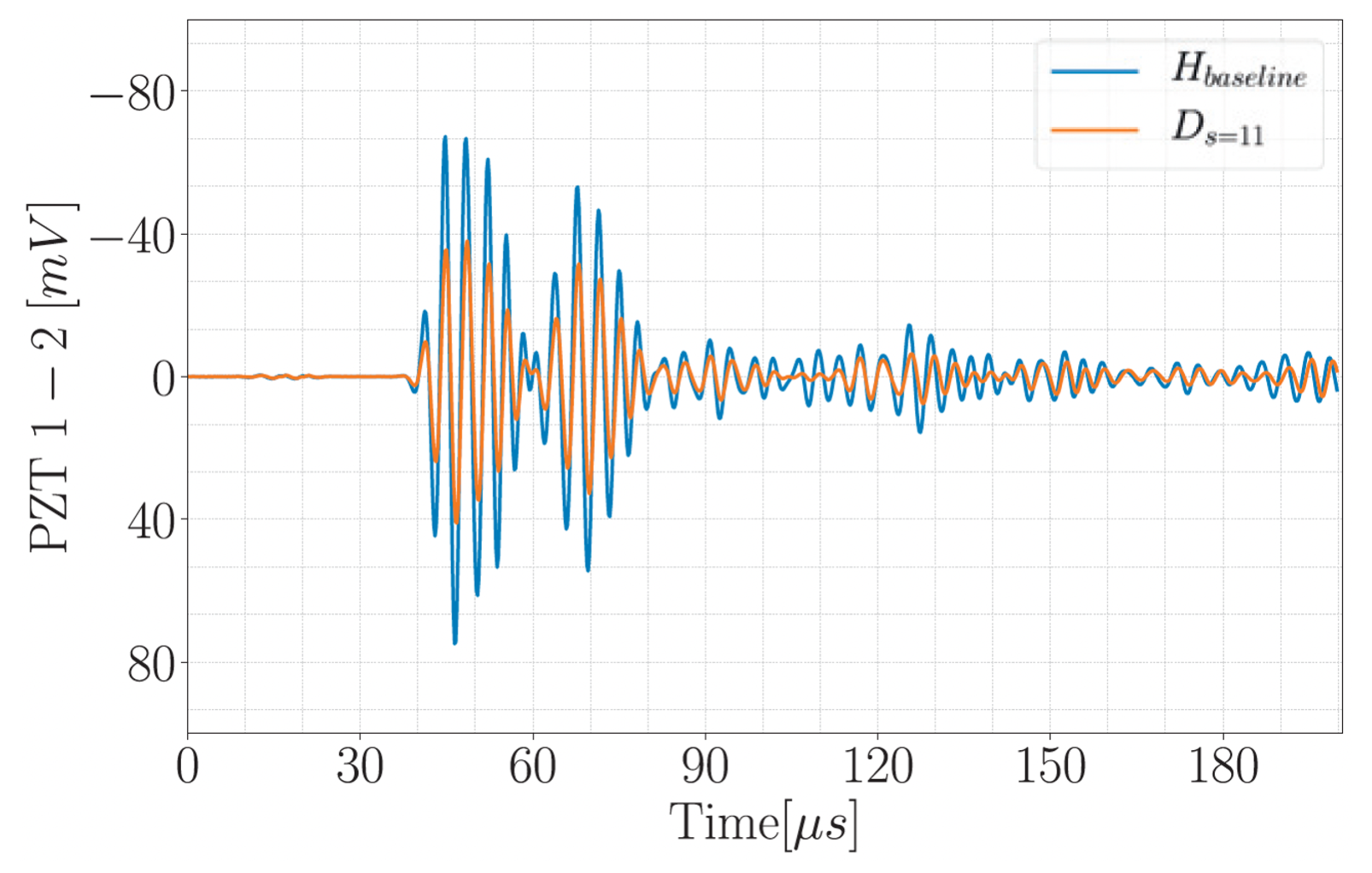}
    \caption{Composite plate and the experimental setup used for generating UNESP-CONCEPT dataset \cite{paixao2021delamination}: (upper) schematic view of the experimental setup (dimensions in mm), and (lower) measurements at PZT 2 sensor location under healthy (blue) and damaged (orange) condition.}
    \label{fig:exp_data}
    \vspace{-1.0em}
\end{figure}

The field of Structural Health Monitoring (SHM) is appropriately viewed as a system identification problem whereby the state of structural damage is inferred from measured structural vibrations.
To demonstrate the effectiveness of the proposed system identification method in such contexts, an experiment was conducted using vibrational data from the UNESP-CONCEPT dataset.  \cite{da2021extrapolation, paixao2021delamination, 10.1007/978-981-13-8331-1_63, Silva2018}. This dataset comprises vibrational data obtained from a carbon-epoxy laminate plate in both its healthy and damaged conditions. Damage was introduced by applying an industrial adhesive putty to the structure, and the extent of damage was systematically adjusted by varying the coverage area of the adhesive putty. The plate was equipped with four piezoelectric transducers (PZT) for data collection. PZT 1 was employed to induce vibrations in the structure, while the remaining three sensors (PZT 2-4) were utilized to measure data. The upper panel of Fig. \ref{fig:exp_data} provides a schematic representation of the experimental setup, while the lower panel illustrates a sample signal acquired under both the baseline (healthy) and damaged conditions.

We carried out an experiment to distinguish between different degrees of damage in a composite plate and its healthy state. We used data from three sensors (PZT 2, PZT 3, and PZT 4) separately for this task. The aforementioned dataset contains measurements for eleven distinct damage levels, where the extent of damage was simulated by varying the area of the adhesive putty. Consequently, this problem evolved into a twelve-class classification task, with one class representing the healthy condition. First, we split the PZT 2 sensor measurements into training and testing sets, with approximately 50 samples per class for both training and testing. Then, we used this data to train and evaluate the classification models described earlier. The second column in Table \ref{table:damage_det} displays the classification accuracy of the comparative methods when using data from PZT 2. Subsequently, we repeated the same experiment for the sensor measurements obtained from PZT 3 and PZT 4 separately. Columns 3 and 4 in the table present the classification performance of the comparative methods using data from PZT 3 and PZT 4, respectively. The results clearly indicate that our proposed method outperformed the other approaches in accurately identifying different extents of structural damage.

\begin{table}[!htb]
    \centering
    \caption{Detection accuracy ($\%$) for different levels of damage using measurements from three sensors, separately. Number of classes: 12, where one class represents a healthy condition, and the remaining 11 classes represent different damage levels.. All methods were trained using $\sim 50$ samples per class and tested on $\sim 50$ test samples per class.}
    {\small
    \begin{tabular}{|l|c|c|c|}
    \hline
        Methods & PZT 2 & PZT 3 & PZT 4 \\
        \hline
        MLP & $49.07$ & $66.47$ & $68.47$ \\
        1D-VGG & $47.20$ & $36.9$ & $26.27$ \\
        1D-ResNet & $10.17$ & $11.93$ & $9.87$ \\
        LSTM-FCN & $9.23$ & $18.47$ & $24.40$ \\
        SVM & $19.33$ & $14.33$ & $14.33$ \\
        FT-SVM & $\mathbf{100.0}$ & $46.0$ & $48.67$ \\
        SCDT-NS & $\mathbf{100.0}$ & $91.83$ & $93.67$ \\
        \textbf{SCDT-NLS} & $\mathbf{100.0}$ & $\mathbf{99.67}$ & $\mathbf{98.33}$ \\
        \hline
    \end{tabular}
    }
    \label{table:damage_det}
\end{table}


\section{Discussion}
The findings presented in Tables \ref{table:nl_detect_acc} to \ref{table:damage_det} demonstrate that the SCDT-NLS classifier-based approach is highly effective for addressing system identification challenges in wave propagation phenomena. In this paper, we approached the system identification problem as a coarse regression problem. From the results it is evident that the proposed solution outperforms state-of-the-art neural network-based time series classification techniques as well as traditional approaches. Furthermore, Table \ref{table:damage_det} illustrates the superior performance of the proposed method in the detection of structural damage when compared to alternative methods.

The proposed solution utilizes sensor data from a specific location to recover the system parameter of interest. This approach operates under the assumption that the sensor data conforms to the signal class model presented in equation (\ref{eq:gen_mod0}). Meaning, for a given system parameter (e.g., $\beta$, $M$, and so on), the signal measured by sensor at a particular location can be viewed as an instance of a template observed under a smooth invertible warping function. However, the proposed system identification technique does not require knowledge of the template or the warping function to recover the parameter of interest. If the generation of an unknown signal satisfies the signal class model, the proposed approach can detect and estimate the system parameter by searching for the nearest local subspace in the SCDT domain using a set of training samples.

In case of a dynamical system where the PDE model can be analytically formulated, the training samples can be obtained by employing a PDE simulation technique, such as the spectral or finite difference method. Conversely, if the PDE model is unknown, training samples can be generated through a controlled experiment with known parameter values. However, generating a large number of training samples under certain conditions (e.g., damaged medium, dispersion, etc.) can be challenging in many structural health monitoring applications. As depicted in Fig.~\ref{fig:est_beta10}, the proposed approach provides superior estimation accuracy with fewer training samples compared to alternative methods. This feature renders the proposed solution suitable for structural health monitoring applications where generating numerous sensor data is often impractical. 

To summarize, this paper proposes a novel approach to address system identification problems pertaining to PDE models of physical phenomena by leveraging sensor data. Specifically, the signals measured at a specific sensor location are considered as observations of a set of template patterns subjected to some unknown time-warpings. While certain PDE systems, like the classical wave equation or convection-diffusion equation, have analytical formulations for sensor measurement that align with this model, others pose difficulties in obtaining such solutions. However, through comprehensive experimentation, this study shows that the proposed signal class model is an effective way to model sensor measurements even in complex PDE systems.

\section{Conclusion} \label{sec:conclusion}
This paper investigated the viability of employing a mathematical model-based approach to address system identification challenges. In the system investigated here, it is presumed that changes occurring in the propagation medium (such as damage, cracks, temperature variations, etc.) of a dynamical system lead to alterations in the parameters of the underlying PDE model. 
Under the assumption that the sensor data adheres to a specific signal class model, we formulated the system identification as a coarse regression problem and employed the SCDT nearest local subspace classifier to detect and estimate the system parameter of interest for a given dynamical system. Extensive numerical experimentation was conducted, and the proposed solution was found to provide a significantly better estimate of the system parameter when compared to state-of-the-art data-driven pattern recognition methods. Moreover, through experimentation using actual vibrational data, the proposed method demonstrated its superiority over the alternative methods in identifying structural damage.

\section*{Acknowledgements}
This work was supported in part by Office of Naval Research (ONR) grant N000142212505, and National Institutes of Health (NIH) grant GM130825. The authors would also like to thank Structural Health Monitoring Research Group (SHM Lab) at São Paulo State University (UNESP) for the UNESP-CONCEPT dataset.

\appendix

\section{Signal Class Model for Diffusion Equation} \label{app:diff}
The proposed signal class model can be applied to the diffusion equation. The 1D diffusion equation is given by,
\begin{equation}
    \dot{u}(x,t) - Du_{xx}(x,t) = 0,
    \label{eq:diff_eq}
\end{equation}
where $D$ represents the diffusion coefficient. Given, the initial condition $u(x,0)=\frac{1}{\sqrt{4\pi}}e^{-\frac{x^2}{4}}$, we can obtain the solution to Eq.~(\ref{eq:diff_eq}) as:
\begin{equation}
    u(x,t)=\frac{1}{\sqrt{4\pi Dt}}e^{-\frac{x^2}{4Dt}},
    \label{eq:sol_diff_eq}
\end{equation}
from which, we can derive the expression for the sensor measurement at $x=x_m$, i.e., $s(t) = \frac{1}{\sqrt{4\pi Dt}}e^{-\frac{x_m^2}{4Dt}}$. Similar to the wave equation, we can find a template $\varphi_D$ and a warping function $g\in\mathcal{G}_D$, both parameterized with $D$, for the diffusion equation so that $s(t)$ can be represented using the mathematical model given as follows:
\begin{align}
    &s(t) = \dot{g}_D(t)\varphi_D(g_D(t)),\quad g_D\in\mathcal{G}_D \nonumber\\
    &\text{template: }\varphi_D(t) = \frac{x_m}{D\sqrt{4\pi t}}e^{-\frac{1}{4t}} \nonumber\\
    &\text{time-warping: }g_{D}(t) = \frac{Dt}{x_m^2}.
    \label{eq:gen_mod_DE}
\end{align}
The inverse of the warping function is given by $g_D^{-1}(t) = \frac{x_m^2t}{D}$, which generates a family of functions $\mathcal{G}^{-1}_D$ that is convex with respect to $\frac{1}{D}$. Hence, the sensor measurements for the diffusion equation also adhere to the proposed signal class model.

\section{Signal Class Model for Convection-Diffusion Equation} \label{app:conv-diff}

1D convection-diffusion equation (PDE) initialized with the initial condition $u(x,0)$ is given by:
\begin{align} \label{eq:conv_diffusion_app}
 &\dot{u}(x,t) = \nu u_x(x,t) + D u_{xx}(x,t),\quad x \in \Omega_x \,, t \in \mathbb R_+, \\
 &\text{Initial condition: }u(x,0) = \frac{1}{\sqrt{4 \pi}} e^{-\frac{x^2}{4}},\nonumber
\end{align}
where $\nu$ and $D$ denote the wave speed and the diffusion coefficient, respectively. 
A solution to the PDE defined in Eq.~(\ref{eq:conv_diffusion_app}) can be derived as:
\begin{align}
    u(x,t) = \frac{1}{\sqrt{4\pi Dt}}e^{-\frac{(x-\nu t)^2}{4Dt}}.
\end{align}
The expression for the sensor data $s(t)$ at location $x=x_m$ is given by:
\begin{align}
    s(t) = \frac{1}{\sqrt{4\pi Dt}}e^{-\frac{(x_m - \nu t)^2}{4Dt}},
    \label{eq:sensor_convdiff_app}
\end{align}
which can be represented using the mathematical model:
\begin{align}
    s(t) &= \dot{g}_{\nu,D}(t)\varphi_{\nu,D}(g_{\nu,D}(t)),\quad g_{\nu,D}\in\mathcal{G}_{\nu,D}, 
    \label{eq:convdiff_genmod_app}
\end{align}
where an inhomogeneous template $\varphi_{\nu,D}(x \,, t) \,:\, \Omega_x \times \mathbb R_+ \rightarrow \mathbb R$ and a warping function in time are:
\begin{align}\label{eq:convdiff_genmod_temp_map_app}
    \varphi_{\nu,D}(t) &= f_{x_m}(t) e^{-t},\\
    g_{\nu,D}(t) &= \frac{(x_m - \nu t)^2}{4 D t},\nonumber\\
    &x_m,\nu,D>0,\quad t > \frac{x_m}{\nu}.\nonumber
\end{align}
Note that the condition $t > \frac{x_m}{\nu}$ ensures $g_{\nu,D}(t)$ to be monotonically increasing, i.e., $\dot{g}_{\nu,D}(t)>0$. Then, we can choose a support of time $t$ as:
\begin{align}  \label{eq:support:t}
 \text{sup}(t) := \left[ t_0 = \frac{x_m}{\nu} + a \,, t_1 \right] := \Omega_t \,,
\end{align}
for $a > 0$ and a large enough $t_1 > t_0$. Here, $f_{x_m}(t)$ is an arbitrary function which will be derived later. First, we derive the inverse of the warping map, i.e., $g^{-1}_{\nu,D}(t)$. To distinguish between the warping map and its inverse, we replace the independent variable $t$ in $g^{-1}_{\nu,D}(t)$ with another variable $z$. Meaning, $g_{\nu,D}^{-1}(z)$ is defined by letting:
\begin{align}  \label{eq:inversewarp:quaraticequation}
 z &= g_{\nu,D}(t) = \frac{(x_m - \nu t)^2}{4 D t}
 \\  
 \label{eq:quaraticequation}
 \Leftrightarrow~&
 \nu^2 t^2 - 2 \left( x_m \nu + 2 D z \right) t + x_m^2 = 0
\end{align}
A condition of existence for solution is: 
$$
\Delta := \left( x_m \nu + 2 D z \right)^2 - \nu^2 x_m^2
= 4 \left( D^2 z^2 + \nu D x_m z \right) > 0 \,. 
$$
Since $\nu > 0$, we have $\Delta > 0$, i.e. equation (\ref{eq:quaraticequation}) always have two solutions.
Note that if $\nu < 0$, then for $\nu < - \frac{x_m}{t}$, the condition becomes (for $x_m \,, z > 0$):
$$
\Delta > 0
~\Leftrightarrow~
- \frac{ D z }{ x_m } < \nu < - \frac{x_m}{t} \,.
$$
In conclusion, this quadratic form has 2 solutions:
\begin{align}  \label{eq:quadraticsolutionsolution}
 t_{\pm} := \frac{1}{\nu^2} \left( x_m \nu + 2 D z 
 \pm \sqrt{ \left( x_m \nu + 2 D z \right)^2 - \nu^2 x_m^2 }
 \right) \,.
\end{align}
Under the condition that $g_{\nu,D}$ and $g^{-1}_{\nu,D}$ are one-one increasing functions, we choose the inverse of the warping map to be,
\begin{align} \label{eq:inverwarpingmap:convecdiff_app}
 g_{\nu,D}^{-1}( z) &=
 \frac{1}{\nu^2} \left( x_m \nu + 2 D z 
 + \sqrt{ \left( x_m \nu + 2 D z \right)^2 - \nu^2 x_m^2 }
 \right) \,.
\end{align}
Eq.~(\ref{eq:inverwarpingmap:convecdiff_app}) can be written as a function $t$ instead of $z$ such that
\begin{align} \label{eq:inverwarpingmap:convecdiff1_app}
 g_{\nu,D}^{-1}( t) &=
 \frac{1}{\nu^2} \left( x_m \nu + 2 D t
 + \sqrt{ \left( x_m \nu + 2 D t \right)^2 - \nu^2 x_m^2 }
 \right) \,.
\end{align}

Let, the family of the functions $g^{-1}_{\nu,D}$ forms a set $\mathcal{G}^{-1}_{\nu,D}$ which is non-convex. However, in section \ref{sec:convdiff}, we demonstrated that the set $\mathcal{G}^{-1}_{\nu,D}$ can be approximated as a convex set with respect to $D$. Here, we aim to derive a similar approximation of $\mathcal{G}^{-1}_{\nu,D}$ which is convex with respect to the parameter $\nu$.

\subsection{Convex approximation of set of inverses of warping maps:} \label{app:conv-diff_convex}
We aim to convexify the set $\mathcal{G}^{-1}_{\nu,D}$ by taking the 1st order Taylor expansion of the function $g_{\nu,D}^{-1}(t)$ with respect to $D$ at $D_0$ (since, $D\in [D_0-\epsilon, D_0+\epsilon]$):
\begin{align*}  
 g_{\nu,D}^{-1}(t) \approx &g_{\nu,D_0}^{-1}(t) 
 + \frac{\partial g_{\nu,D}^{-1}(t)}{\partial D} \Big |_{D = D_0} (D - D_0)
 \notag
 \\  
 = &\frac{1}{\nu^2} \left( x_m \nu + 2 D_0 t + \sqrt{ (x_m \nu + 2 D_0 t)^2 - \nu^2 x_m^2 } \right)\\
 &+ \frac{2 t}{\nu^2} \left( 1 + \frac{ x_m \nu + 2 D_0 t }{ \sqrt{ (x_m \nu + 2 D_0 t)^2 - \nu^2 x_m^2 } } \right) (D - D_0)
 \\  
 := &\tilde{g}_{\nu,D}^{-1}(t) \,.
\end{align*}
Since $D \in \left[ D_0 - \epsilon \,, D_0 + \epsilon \right]$, we evaluate the approximate inverse warping map $\tilde{g}_{\nu,D}^{-1}(t)$ at the boundary values of $D$, i.e., $D = D_0 \pm \epsilon$ to obtain:
\begin{align*}  
 g_{\nu,D_0 \pm \epsilon}^{-1}(t) \approx \tilde{g}_{\nu,D_0 \pm \epsilon}^{-1}(t) 
  = &\frac{1}{\nu^2} \left( x_m \nu + 2 D_0 t + \sqrt{ (x_m \nu + 2 D_0 t)^2 - \nu^2 x_m^2 } \right)\\
  &\pm \frac{2 t}{\nu^2} \left( 1 + \frac{ x_m \nu + 2 D_0 t }{ \sqrt{ (x_m \nu + 2 D_0 t)^2 - \nu^2 x_m^2 } } \right) \epsilon
 \\  
 = &g_{\nu,D_0}^{-1}(t)
 \pm \frac{2 t}{\nu^2} \left( 1 + \frac{ x_m \nu + 2 D_0 t }{ \sqrt{ (x_m \nu + 2 D_0 t)^2 - \nu^2 x_m^2 } } \right) \epsilon.
\end{align*}
The expression for the approximated inverse warping map above suggests that the set $\mathcal{G}_{\nu,D}^{-1}$ can be approximated to be convex regarding the parameter $D$.

Similarly, we can also convexify the set $\mathcal{G}^{-1}_{\nu,D}$ by taking the 1st order Taylor expansion of the function $g^{-1}_{\nu,D}$ with respect to $\nu$ at $\nu_0$ (since, we assume $\nu \in [\nu_0 - \epsilon \,, \nu_0 + \epsilon ]$):
\begin{align} 
 &g_{\nu,D}^{-1}(t) \approx g_{\nu_0,D}^{-1}( t) 
 + \frac{\partial g_{\nu,D}^{-1}(t)}{\partial \nu} \Big |_{\nu = \nu_0} (\nu - \nu_0)
 \notag
 \\  
 &= \frac{1}{\nu_0^2} \left( x_m \nu_0 + 2 D t 
 + \sqrt{ \left( x_m \nu_0 + 2 D t \right)^2 - \nu_0^2 x_m^2 }
 \right) 
 \notag
 \\&  
 - \frac{1}{\nu_0^2} \left( x_m + \frac{4 D t}{\nu_0}  
 + \frac{2}{\nu_0} \sqrt{ \left( x_m \nu_0 + 2 D t \right)^2 - \nu_0^2 x^2 }
 - \frac{ 2 D t x_m }{ \sqrt{ \left( x \nu_0 + 2 D t \right)^2 - \nu_0^2 x_m^2 } } \right) (\nu - \nu_0)
 \notag 
 \\ 
 &:= \tilde{g}_{\nu,D}^{-1}(t) \,. \notag
\end{align}
We then evaluate the $\tilde{g}_{\nu,D}^{-1}(t)$ at the boundary values of $\nu$, i.e., $\nu = \nu_0 \pm \epsilon$ to obtain:
\begin{align}
    &g_{\nu_0\pm\epsilon,D}^{-1}(t) \approx \tilde{g}_{\nu_0\pm\epsilon,D}^{-1}(t)\notag \\
    &= \frac{1}{\nu_0^2} \left( x_m \nu_0 + 2 D t 
 + \sqrt{ \left( x_m \nu_0 + 2 D t \right)^2 - \nu_0^2 x_m^2 }
 \right) 
 \notag
 \\&  
 \pm \frac{1}{\nu_0^2} \left( x_m + \frac{4 D t}{\nu_0}  
 + \frac{2}{\nu_0} \sqrt{ \left( x_m \nu_0 + 2 D t \right)^2 - \nu_0^2 x^2 }
 - \frac{ 2 D t x_m }{ \sqrt{ \left( x \nu_0 + 2 D t \right)^2 - \nu_0^2 x_m^2 } } \right) \epsilon \notag\\
 &= g_{\nu_0,D}^{-1}(t)\notag \\ 
 &\pm \frac{1}{\nu_0^2} \left( x_m + \frac{4 D t}{\nu_0}  
 + \frac{2}{\nu_0} \sqrt{ \left( x_m \nu_0 + 2 D t \right)^2 - \nu_0^2 x^2 }
 - \frac{ 2 D t x_m }{ \sqrt{ \left( x \nu_0 + 2 D t \right)^2 - \nu_0^2 x_m^2 } } \right) \epsilon, \notag
\end{align}
which is convex with respect to $\epsilon$. It indicates that the set $\mathcal{G}^{-1}_{\nu,D}$ is convex with respect to $\nu \in [\nu_0-\epsilon, \nu_0+\epsilon]$ for a given $D$.

\subsection{Derivation of template} To derive the template $\varphi_{\nu,D}(t)$ defined in Eq.~(\ref{eq:convdiff_genmod_temp_map_app}), we need to find an expression for the function $f_{x_m}(t)$. We choose $f_{x_m}(t)$ such that:
\begin{align*}  
 \frac{1}{4 D} 
 \left( \nu^2 - \frac{x_m^2}{t^2} \right) 
 f_{x_m} \left( \frac{(x_m - \nu t)^2}{4 D t} \right) 
 e^{  - \frac{(x_m - \nu t)^2}{4 D t} } = \frac{1}{\sqrt{4 \pi D t}} e^{ - \frac{(x_m - \nu t)^2}{4 D t} } \,.
\end{align*}
The above equality holds if we choose $f_{x_m}(t)$ in such a way that:
\begin{align*}
 &\frac{1}{\sqrt{4 \pi D t}} = \frac{1}{4 D} 
 \left( \nu^2 - \frac{x_m^2}{t^2} \right) 
 f_{x_m} \left( \frac{(x_m - \nu t)^2}{4 D t} \right) 
 \\  
 \Leftrightarrow~&
 f_{x_m} \left( \frac{(x_m - \nu t)^2}{4 D t} \right)
 = \sqrt{ \frac{4 D}{\pi} }
 \frac{t^{\frac{3}{2}}}{\nu^2 t^2 - x_m^2}  
 \notag
 \\  
 \Leftrightarrow~&
 f_{x_m} \left( t \right) = \frac{1}{\nu} \sqrt{ \frac{4 D}{\pi} }
 \frac{ \left( \left( x_m \nu + 2 D t \right)
 + \sqrt{ \left( x_m \nu + 2 D t \right)^2 - \nu^2 x_m^2 }
 \right)^{\frac{3}{2}} }
 { \left( \left( x_m \nu + 2 D t \right)
 + \sqrt{ \left( x_m \nu + 2 D t \right)^2 - \nu^2 x_m^2 }
 \right)^2 - \nu^2 x_m^2 } \,.
\end{align*}
Thus, the template is given by:
\begin{align}  \label{eq:convectiondiffusiontime:template:final}
 \varphi_{\nu,D}(t) &= \frac{1}{\nu} \sqrt{ \frac{4 D}{\pi} }
 \frac{ \left( \left( x_m \nu + 2 D t \right)
 + \sqrt{ \left( x_m \nu + 2 D t \right)^2 - \nu^2 x_m^2 }
 \right)^{\frac{3}{2}} }
 { \left( \left( x_m \nu + 2 D t \right)
 + \sqrt{ \left( x_m \nu + 2 D t \right)^2 - \nu^2 x_m^2 }
 \right)^2 - \nu^2 x_m^2 } 
 e^{-t} \,.
\end{align}

\section{Nondimensionalization Procedure}\label{app:nondim}

	\newcommand{\xnd}{X}
	\newcommand{\tnd}{T}
	\newcommand{\und}{U}
	\newcommand{\xc}{x_0}
	\newcommand{\wc}{\omega_0}
	\newcommand{\uc}{u_0}
	In this section, we obtain a dimensionless version of the PDE defined by  Eq.~\eqref{eq:we_damaged}, which we repeat here for convenience:
	\begin{equation}\label{eq:we_damaged_App}
		\rho \ddot{u}-Eu_{xx}-\eta \dot{u}_{xx}-\rho M \ddot{u}_{xx} - Fu_{xxxx} + \mathcal{B} u_x u_{xx} = \Phi.
	\end{equation}
	We begin by introducing the dimensionless variables
	\begin{subequations}\label{eq:nondimvars}
		\begin{align}
			\xnd &= \frac{x}{\xc},\\
			\tnd &= \wc t, \\
			\und &= \frac{u}{\uc},
		\end{align}
	\end{subequations}
	where $ \xc $, $ \wc $, and $ \uc $ are positive constants defining the characteristic length, frequency, and displacement, respectively, for propagating waves in the system. Substituting the change of variables defined by Eqs.~\eqref{eq:nondimvars} into Eq.~\eqref{eq:we_damaged_App}, and dividing the result by the quantity $ \rho \uc \wc^2 $, yields the following dimensionless PDE:
	\begin{equation}\label{key}
		\xi_0 \ddot{U} - \xi_1 U_{\xnd \xnd} - \xi_2 \dot{U}_{\xnd \xnd} - \xi_3 \ddot{U}_{\xnd \xnd} - \xi_4 U_{\xnd \xnd \xnd \xnd} + \beta U_\xnd U_{\xnd \xnd} = \phi,
	\end{equation}
	where a prime denotes partial differentiation with respect to $ \tnd $, and the dimensionless coefficients and force are given by
	\begin{subequations}\label{eq:nondimcoeff}
		\begin{align}
			\xi_0 &= 1,\\
			\xi_1 &= \frac{E}{\rho \wc^2 \xc^2}, \\
			\xi_2 &= \frac{\eta}{\rho \wc \xc^2}, \\
			\xi_3 &= \frac{M}{\xc^2} \\
			\xi_4 &= \frac{F}{\rho \wc^2 \xc^4}, \\
			\beta &= \frac{\mathcal{B} \uc}{\rho \wc^2 \xc^3}, \\
			\phi &= \frac{\Phi}{\rho \uc \wc^2}.
		\end{align}
	\end{subequations}
	In the main text (i.e., beginning with Eq.~\eqref{eq:we_gen}), we retain the variable names of Eq.~\eqref{eq:we_damaged_App} and implicitly consider dimensionless variables thereafter.
	
	A suitable choice of the scales, $ \xc $, $ \wc $, and $ \uc $, requires some physical knowledge of a given system. However, assuming the distortion mechanisms can be viewed as perturbations to d'Alembertian traveling waves (i.e., the overall behavior remains wave-like and is not dominated by damping, dispersion, or nonlinearity), the following procedure should yield a dimensionless PDE with coefficients that are at most $ O(1) $:
	\begin{enumerate}
		\item Assume that estimates of the mass density $ \rho $ and characteristic wave speed $ \sqrt{E/\rho} $ are available with reasonable uncertainty, e.g., by performing simple measurements on undamaged material samples.
		\item Set the characteristic length $ \xc $ equal to the largest wavelength at which dispersive effects (i.e., the terms proportional to $ \xi_3 $ and $ \xi_4 $) are expected to be significant. As an example, for a thin, elastic rod this could correspond to the diameter of the rod or the spatial period of a periodic microstructure.
		\item Choose the characteristic displacement $ \uc $ such that the strain-like quantity $ \uc/\xc $ is equal to the largest strain expected under experimental conditions.
		\item Choose the characteristic frequency $ \wc $ such that the coefficient $ \xi_1 = 1 $ when the variables $ \rho $, $ E $, and $ \xc $ take the values defined above.
	\end{enumerate}
	Under the assumptions listed above, with these scalings, the resulting dimensionless PDE has a characteristic wave speed given by $ \sqrt{\xi_1} \approx 1 $, with all other coefficients taking values $ O(1) $ or smaller.


\bibliographystyle{elsarticle-num} 
\bibliography{reference}

\end{document}